# Universal programmable and self-configuring optical filter

David A. B. Miller*[,1], Charles Roques-Carmes[1], Carson G. Valdez[1], Anne R. Kroo[1], Marek Vlk[1,2], Shanhui Fan[1] and Olav Solgaard[1]

[1]*Stanford University, Ginzton Laboratory, 348 Via Pueblo Mall, Stanford CA 94305*
[2]*Department of Physics and Technology, UiT The Arctic University of Norway, NO-9037 Tromsø, Norway*
**dabm@stanford.edu*

**Abstract:** We propose an approach to integrated optical spectral filtering that allows arbitrary programmability, can compensate automatically for imperfections in filter fabrication, allows multiple simultaneous and separately programmable filter functions on the same input, and can configure itself automatically to the problem of interest, for example to filter or reject multiple arbitrarily chosen frequencies. The approach exploits splitting the input light into an array of multiple waveguides of different lengths that then feed a programmable interferometer array that can also self-configure. It can give spectral response similar to arrayed waveguide gratings but offers many other filtering functions, as well as supporting other structures based on non-redundant arrays for precise spectral filtering. Simultaneous filtering also allows, for the first time to our knowledge, an automatic measurement of the temporal coherency matrix and physical separation into the Karhunen-Loève expansion of temporally partially coherent light fields.

## 1. Introduction

Optical frequency or wavelength filters [1] are used in a wide variety of applications, including wavelength-division multiplexing in telecommunications and spectroscopy for sensing – for example, of different chemicals or gases. Existing filters and spectrometers include those based on gratings or prisms, multilayer dielectric stacks, resonators, or arrayed waveguide gratings (AWG) [2–8]. Many such filters have little or no programmability after manufacture. Some allow tuning by mechanical movement, e.g., of a grating in a spectrometer. Waveguide array and resonator approaches can be made in integrated photonic circuits, which can allow thermal, optoelectronic, micromechanical or piezoelectric tuning and other adjustments. Such circuits based on meshes of interferometers can be very programmable [9,10]; "recirculating" interferometer mesh architectures show impressive programmable filtering based on resonating rings [11]. The other major category of interferometer meshes – "forward only" architectures [9,12–15] – allows simple progressive programming and even self-configuration [9,12–14,16–20], but so far only for manipulating spatial fields and modes. Now we show how such forward-only meshes can perform programmable and self-configuring spectral filtering as well as novel measurement and separation of partially coherent light.

Our approach exploits the combination of an array of waveguides of different lengths (as in an AWG) feeding a forward-only interferometer mesh. This opens a wide range of novel and programmable spectral capabilities, some without precedent in conventional filters. Our approach can implement arbitrary filters in a broad set of possibilities; these include multiple-layer filters that give multiple separate programmable filter functions at the same time (so also allowing switching of wavelength channels between outputs). Novel applications include measuring temporal coherence and separating temporally partially coherent fields to their mutually incoherent components. Self-configuration gives automatic tuning to incident wavelengths and allows automatic compensation for fabrication imperfections, such as imperfect waveguide lengths. Such compensation opens new architectures including non-redundant array waveguide lengths for precise filtering.

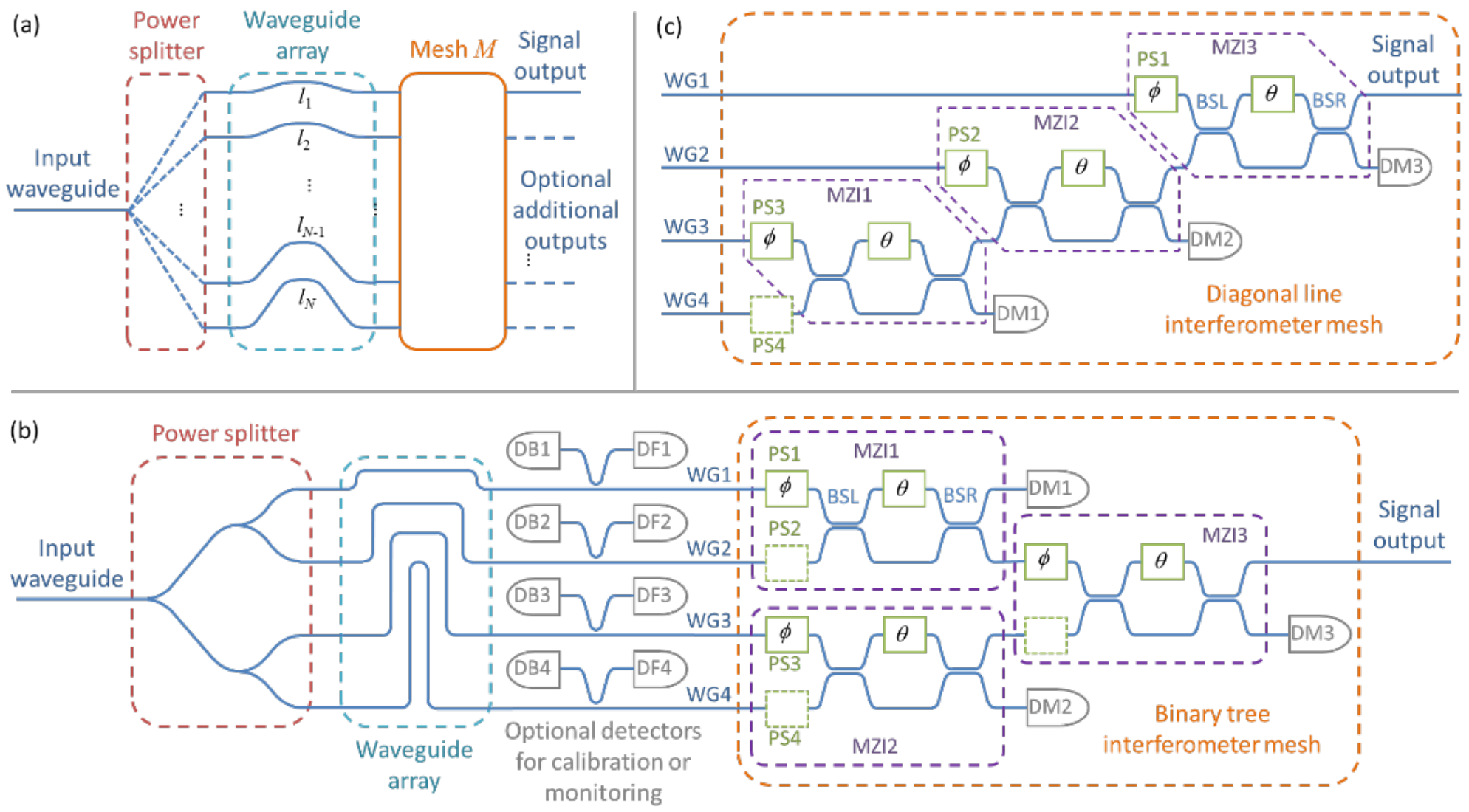

Fig. 1. (a) Block diagram with power splitter, waveguide array, and interferometer mesh. (b) Example with a (symmetric) binary tree single-layer interferometer mesh [12,14] with 4 waveguides. (c) Alternative diagonal line single-layer mesh [12,14]. Square box elements are controllable phase shifters, with at least two required, one "$\phi$" and one "$\theta$", for each Mach-Zehnder interferometer (MZI). (The phase shift values in the "$\phi$" and "$\theta$" shifters in different MZIs will generally be different.) Each MZI also requires two beam splitters, BSL and BSR, that ideally have a 50:50 split ratio. Detectors DM1 – DM3 can be used to configure the mesh. Optional "tap" detectors DB1 – DB3 and DF1 – DF3, based on sampling a small amount of the waveguide power, can be used in calibration [14] or for monitoring. Optical phase shifters in dashed boxes are not formally required for full programmability, but PS2 and PS4 in (b) and PS4 in (c) could be convenient, e.g., in tuning the "center" frequency of the filter.

## 2. Device concept

Fig. 1(a) illustrates the concept. Light in a single input waveguide is power-split – for example, equally – into an array of multiple waveguides. These waveguides have different lengths $l_p$ – for example, each longer than the preceding one by some specific amount $\delta l_o$. The outputs of those waveguides then feed a programmable "forward-only" interferometer mesh. The power splitter and waveguide array can be similar to those of an AWG router, but our approach differs from some previous partially programmable AWG approaches [8] by using a programmable interferometer mesh at the output.

Suppose, for illustration, that the light in the input waveguide is at one specific wavelength or frequency. Because the waveguides are of different lengths, the amplitudes arriving through these different input waveguides (WG1 to WG4 here) to the interferometer mesh will have different phases. (Indeed, if the power splitting is not equal or the waveguides have different losses, the arriving magnitudes may also be different.) So, the mesh then sees a (spatial) mathematical input (column) vector of different complex amplitudes of light at this frequency, with each different vector element appearing as an amplitude in a different one of the mesh input waveguides (WG1 to WG4 here). The phase delays, and hence the vector, will be different for different wavelengths. Essentially, the power splitting and the waveguide array turn different wavelengths of light into different spatial input vectors to the mesh. This key argument allows us to transfer physics and concepts of self-configuring Mach-Zehnder interferometer (MZI) arrays [9,14] from the spatial to the temporal domain.

The simplest category of filters uses a single-layer mesh – i.e., one having just one signal output, as shown in Fig. 1 (b) and (c), implementing a single, programmable filter function at its signal output. The other "drop" outputs from the single-layer meshes of Fig. 1(b) and (c) are shown being dumped into detectors DM1 – DM3.

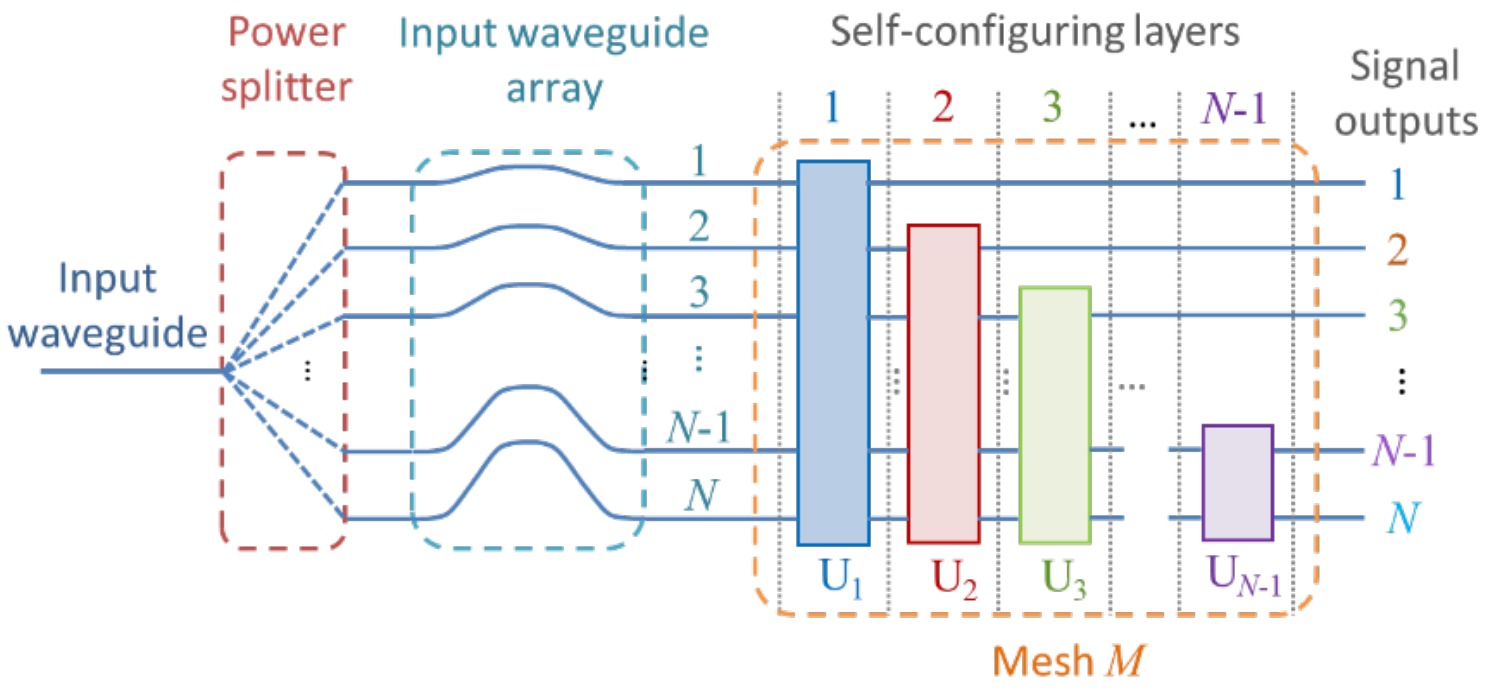

Fig. 2. Device with a mesh $M$ formed from $N-1$ successive self-configuring layers to allow multiple simultaneous filter functions at the various different signal outputs, all operating on the same input light.

Instead of dumping all the power in the "drop" ports of the MZI interferometers (or other equivalent 2-input, 2-output "$2\times2$" blocks [10]) into photodetectors (as shown in Fig. 1(b) and (c)), we could instead pass most or all of this power into a second mesh "layer" (Fig. 2). (The "drop-port" detectors of the first layer – DM1 to DM3 in the example in Fig. 1 – could be made to be mostly transparent [16], we could tap off a small portion of the power to a separate photodetector [21,22], we could otherwise deduce the power in given waveguides [23], or we could use progressive algorithms based on maximizing the power in the signal output only [12].) This allows configuration of a second filter function to appear from the signal output of the second such layer. We can similarly continue to stack further such layers for additional such outputs, up to $N-1$ as in Fig. 2.

Fig. 2 shows the device concept with a complete cascade of layers, in this case for a full set of $N$ programmable output filter functions forming a full arbitrary unitary mesh. In Fig. 2, we represent an entire layer as in Fig. 1(b) or (c) as a vertical rectangle. Though we could use any unitary mesh that allows programming to route an arbitrary input vector to a signal output, the self-configuring layer architecture [9,12,14] is arguably the simplest and most economical universal approach for each such layer; it also allows simple programmability or self-configuration. Each layer $j$ formally implements a unitary transform $U_j$ between its inputs and its outputs, with the entire mesh matrix $M$ being the product of these.

## 3. Device analysis

Quite generally, the mesh of Fig. 1(a) multiplies the vector of amplitudes from the waveguide array by a programmable matrix $M$. Multiple-layer interferometer meshes allow the construction of arbitrary matrices $M$ [9]. Programming this matrix, either by calculated setting of interferometer parameters or by self-configuration to the problem of interest, corresponds to programming the spectral response of the system. Such programmability allows a wide and reconfigurable range of filtering behaviors (see Supplementary sections S1 for a detailed analytic approach, S2 for detailed analysis of a simple filter, and S3 for extended discussion of mesh matrices and filter functions).

For example, we can make each waveguide $p$ of the $N$ waveguides in the array have relative lengths $\delta l_p$ that are integer multiples $m_p$ of a length $\delta l_o$, i.e., $\delta l_p = m_p \delta l_o$. These will have corresponding relative time delays $m_p \delta t_o$, where

$$\delta t_o = n_g \delta l_o / c \tag{1}$$

for waveguide group velocity $n_g$ and free-space velocity of light $c$. (This will lead to a filter with a free-spectral range of $\omega_{FSR} = 2\pi / \delta t_o$ in angular frequency or $f_{FSR} = 1/\delta t_o$ in conventional frequency.)

We can consider an input (angular) frequency $\delta\omega$ relative to some center frequency. So, for an input amplitude $x(\delta\omega)$ as a function of frequency in the input waveguide, for each output waveguide $q$ we can separately program a filter frequency response $H_q(\delta\omega)$ using settings of the mesh matrix $M$. (We presume $M$ itself is essentially independent of frequency over some reasonable frequency range of interest.). The corresponding signal output in waveguide $q$ as a function of frequency, $y_q(\delta\omega)$, is then

$$y_q(\delta\omega) = H_q(\delta\omega)\, x(\delta\omega) \tag{2}$$

The relative phase delay in input waveguide $p$ will be $m_p \delta\omega \delta t_o$, leading to a propagation factor $\exp(-i\, m_p \delta\omega \delta t_o)$ through that guide. We presume the power splitter leads to relative amplitudes $a_p$ in the array waveguides. (These $a_p$ might all be equal, with a value $\sqrt{1/N}$ for the loss-less case.) Then, for programmed mesh matrix elements $M_{qp}$, these filter functions are given by

$$H_q(\delta\omega) = \sum_{p=1}^{N} a_p M_{qp} \exp(-i\, m_p \delta\omega \delta t_o) \tag{3}$$

for the signal output waveguide $q$. Formally, we can view this as an expansion of each desired function $H_q(\delta\omega)$ on the orthogonal basis (normalized over a free-spectral range)

$$h_p(\delta\omega) = \sqrt{\frac{\delta t_o}{2\pi}} \exp(-i m_p \delta\omega \delta t_o) \tag{4}$$

with the required matrix elements $M_{qj}$ evaluated by premultiplying by $h_p^*(\delta\omega)$ and integrating over a free-spectral range, i.e.,

$$M_{qj} \equiv \frac{1}{a_j} \sqrt{\frac{\delta t_o}{2\pi}} \int_{-\pi/\delta t_o}^{\pi/\delta t_o} \exp(i m_j \delta\omega \delta t_o)\, H_q(\delta\omega)\, d(\delta\omega) \tag{5}$$

Equivalently, we are projecting $H_q(\delta\omega)$ onto this orthogonal basis $h_p(\delta\omega)$, a basis that we choose by the design of the lengths of the waveguides in the array. With these definitions

$$\sum_{j=1}^{N} \left| a_j M_{qj} \right|^2 = \delta t_o \int_{-\pi/\delta t_o}^{\pi/\delta t_o} \left| H_q(\delta\omega) \right|^2 d(\delta\omega) \tag{6}$$

Though Eq. (5) is like establishing a Fourier coefficient for a Fourier-series expansion, in general we call $h_p(\delta\omega)$ just a "Fourier-like" basis because the $m_p$ in successive waveguides are not necessarily successive integers and there is only a finite number $N$ of such basis functions. If the $m_p$ are successive integers (as in a typical AWG), then Eq. (3) is like a conventional Fourier series expansion, except only over a finite basis or set of frequencies. (In a conventional Fourier series, the summation as in Eq. (3) would be from $p = -\infty$ to $+\infty$ with $m_p = p$.)

Any filter response $H_q(\delta\omega)$ that is formed from linear superpositions of these Fourier-like basis functions $h_p(\delta\omega)$ is possible. So, a design procedure for a desired filter function $H_q(\delta\omega)$ is (i) evaluate the necessary matrix elements $M_{qj}$ as in Eq. (5), and (ii) deduce the necessary mesh settings to implement these matrix elements.

Note, of course, that only filter functions that can be written as a superposition of these basis functions $h_p(\delta\omega)$ can be exactly implemented by the device, and those basis functions are set by the choice of waveguide lengths. Otherwise, this approach still gives the best approximation (in a "least squares" sense) to the filter design that the device can implement.

## 4. Single-layer filter

One simple and useful mesh is a single self-configuring layer of MZIs [12,14]. This can be made from a (symmetric) "binary tree" as in Fig. 1(b) or a "diagonal line" of interferometers as in Fig. 1(c), or, indeed, hybrids of these two approaches [14]. For coherent light at a specific frequency, such self-configuring layers can be automatically and progressively configured to direct all the power from any given input vector of amplitudes at that wavelength to their one "signal" output. For example, in the binary tree we can automatically configure the first (upper) interferometer MZI1, by adjusting its $\phi$ phase shifter PS1 to minimize the power in its output detector DM1, and then subsequently adjusting its $\theta$ phase shifter to further minimize that same power. With 50:50 beamsplitters in the interferometers, this will result in zero power in the detector DM1. We can simultaneously perform the same kind of power minimization in the second (lower) interferometer MZI2, using detector DM2. Subsequently we can perform a similar power minimization in the third interferometer MZI3 (in a second "column" of interferometers) using detector DM3. The result is that all the input power is routed to the signal output of the mesh. For the "diagonal line", Fig. 1(c), a similar minimization of one MZI after another also routes all power at this wavelength to its signal output. This progressive approach is easily generalized to larger binary trees with more input waveguides and "columns" of interferometers, to larger diagonal lines, or to other self-configuring layer architectures [14].

This progressive configuration by single-parameter power minimizations gives the "self-configuring" name [9,12], but such layers can also be defined topologically [14,15] through the property that each input to such a "layer" connects to the layer's signal output by only one path through the $2 \times 2$ blocks or interferometers in the layer.

Hence, this spectrometer or filter will have configured itself to route all the light at this wavelength to the signal output, making a self-configuring filter for this wavelength. It is also possible to perform self-configuration just by progressively maximizing the power at the signal output [12–14], so multiple embedded detectors are not essential. These "single-layer" self-configuring mesh architectures are equivalent to those of the self-aligning beam coupler [12] for spatial modes.

Note that no calibration of the device is required for this operation. Unlike a simple fixed AWG device [6,24], this approach is tolerant of waveguide lengths not being perfectly correct – small length errors are automatically compensated by the mesh phase shifters – and of having possibly different losses; with 50:50 beamsplitters in the MZIs all the power at that input wavelength that arrives at the ends of the waveguides will still be routed to the signal output. "Perfect" MZI approaches [13,25] allow MZIs with imperfect fabricated beamsplitter ratios to give the same behavior.

Once set up in this way, the precise form of the filter function for other input wavelengths depends on the actual lengths (and possibly losses) of the different waveguides in the array, but this function is easily calculated (see Supplementary section S1). With equal power splitting, equal loss in all waveguides and equally spaced waveguide lengths – what we could call a "simple" filter – the filter function is the same as that of an AWG router from any one input to any one output, with the additional benefits of tolerance of imprecise fabrication and simple self-configuration to any desired "center" wavelength.

Note, too, that, unlike a conventional AWG router, the "center" frequency of this filter can be tuned simply, e.g., by using the phase shifters PS1 to PS4 in Fig. 1. All the phase shifters in such meshes can be calibrated by relatively simple progressive algorithms with some reference input at one known frequency or wavelength, especially if the backward sampling detectors, e.g., DB1 – DB4 in Fig. 1, are included [14]. Such calibration means any specific filter function as in Eq. (3) can also be directly programmed into the mesh. (See Ref. [14] supplementary information for a comprehensive analysis of MZI blocks and programming a self-configuring layer to collect a given vector or, equivalently in the present case, set a specific frequency response $H(\delta\omega)$.)

Fig. 3(a) shows calculated results for the frequency response of a simple filter based on the approach of Fig. 1, but with an array of 16 waveguides with successive waveguides longer than previous one by a length increment of $\delta l_o = 16.51\mu m$ (as in an AWG). These waveguides feed a 16-input self-configuring layer, such as a 16-input symmetric binary tree. (Such a binary tree would have 8 MZIs in a first column, 4 in a second, 2 in a third and 1 in a fourth.) In this example, the filter is set up to collect all of one specific frequency, as in the peak of the $\gamma = 0$ curve of Fig. 3(a), showing a frequency response similar to that of one port of an AWG.

For our example calculations, we presume a silicon strip waveguide 500 nm wide by 220 nm tall, for which we estimate a group index of $n_g = 4.05647$ near a "center" wavelength of 1550 nm. The chosen length increment $\delta l_o = 16.51\mu m$ gives a free-spectral range of 4.4665 THz, slightly larger than the width of the telecommunications C-band. (See Supplementary Section S1 for a detailed discussion.)

Once set up for any given frequency, such a filter can be tuned by adding phase increments to the phase shifters at each mesh input, e.g., to the phase shifters PS1 – PS4 in Fig. 1. With our presumed waveguides whose lengths differ by an integer multiple $m_p$ of the increment $\delta l_o$, we choose a phase "tilt" number $\gamma$ between 0 and $2\pi$ to add a phase delay

$$\delta_p = m_p \gamma \tag{7}$$

to each of the "input" phase shifters, e.g., PS1 – PS4 in Fig. 1. In this way, we can tune the filter over its complete spectral range. (In practice, one would subtract any integer multiples of $2\pi$ from the phase $\delta_p$ to get a physical phase shift within a $2\pi$ range for an actual phase shifter.) Fig. 3(a) shows this tunability with a set of example values for $\gamma$. One simple way of tuning like this would be to heat up the entire set of waveguides uniformly, relying on the temperature dependence of refractive index in the waveguide material; longer waveguides will acquire proportionately longer phase additions $\delta_p$ as needed in Eq. (7).

Note that, typically, the design of such interferometer meshes themselves is such that both path lengths to any points where pairs of beams interfere inside the mesh are essentially equal, as are all total path lengths from any input to any output; as a result, the mesh itself and the corresponding matrix $M$ are nominally non-dispersive except for dispersion in beamsplitters and in wavelength-dependence of phase shifts. Hence, at least for moderate bandwidths, the performance of the device can be dominated by the different lengths of the waveguide elements, allowing simple design. So, we neglect any other dispersion of the mesh in our calculations for simplicity; any such additional known dispersion could be straightforwardly included in a numerical analysis of the device operation; equivalently, we presume the matrix $M$ does not itself depend on wavelength.

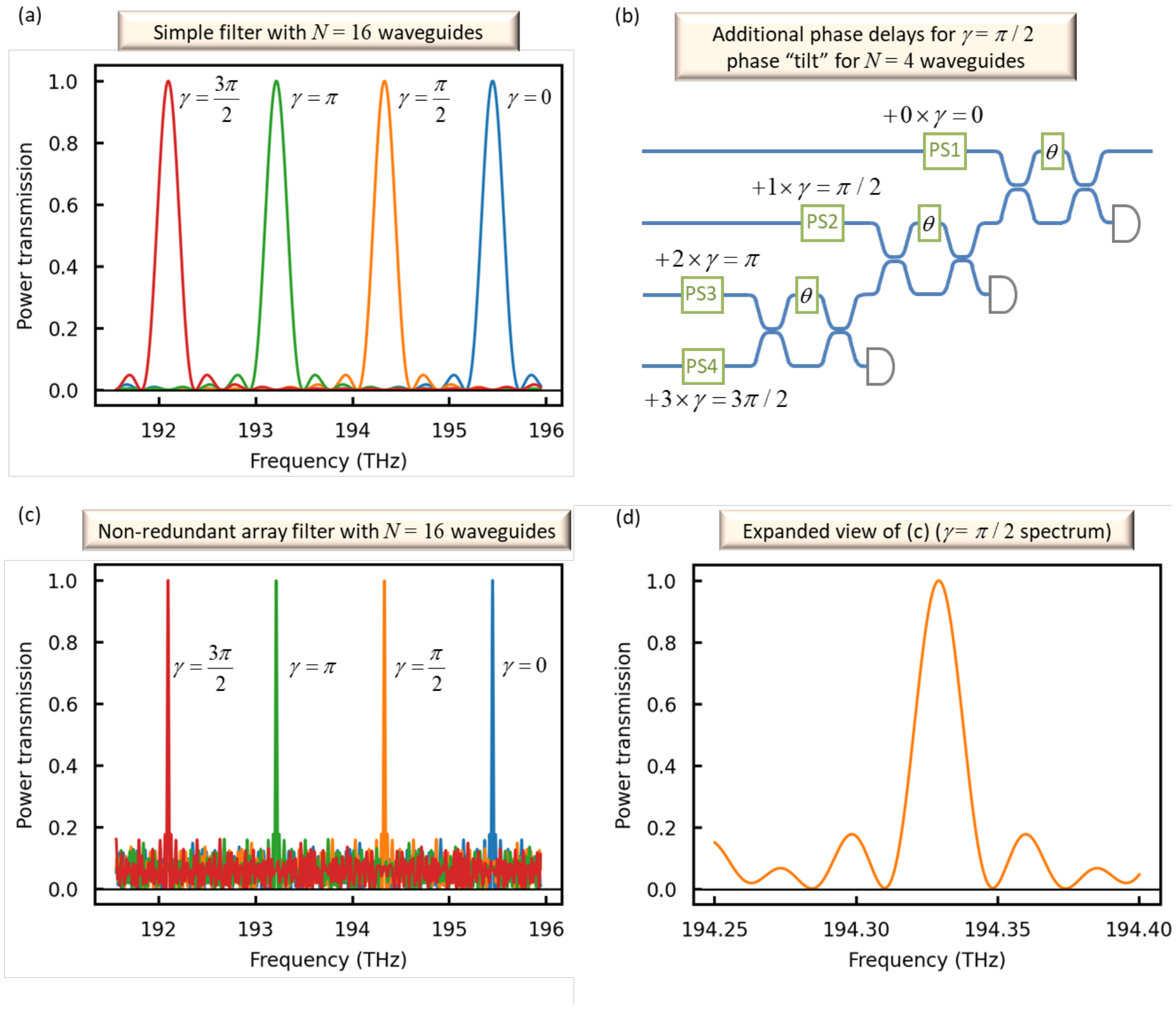

Fig. 3. (a), (c) and (d): Frequency response graphs for a single-layer filter with $N = 16$ waveguides in the array, designed to operate usefully over a frequency range approximately equivalent to the telecommunications C-band (1530 nm to 1565 nm). Waveguide lengths differ by increments of $\delta l_o = 16.51\,\mu\text{m}$. The filter can be tuned by adding a phase "tilt" given by the tuning parameter $\gamma$ times the number $m_p$ of incremental length units $\delta l_o$ for that $p$th waveguide. Four example values of $\gamma$ are shown in (a) and (c). (b) shows the additional phase delays for $\gamma = \pi/2$ for the simpler case of $N = 4$ waveguides (as in the diagonal line mesh of Fig. 1(c) with phase shifters PS1 to PS4). (a) Response for a simple filter where successive waveguides are longer by $\delta l_o$ than the preceding one (as in an AWG). (c) Response for a filter based on a non-redundant array of waveguide lengths that are longer by increments (based on a Golomb ruler) of $[0,1,4,11,26,32,56,68,76,115,117,134,150,163,168,177]\delta l_o$ . (d) The curve for the response of the as in (b) for a tuning parameter of $\gamma = \pi/2$, plotted over a narrower frequency range to show detail of the peak response.

This single-layer filter, especially with constant length (or, equivalently, time-delay) increments between the waveguides in the array, can be viewed as a tapped delay line, transversal, or (non-regressive) moving-average finite-impulse-response filter (see, e.g., [1] pp.177 et seq.), as may be immediately obvious from the "diagonal line" architecture; any relative complex amplitudes of tap weights can be directly programmed into the diagonal line of MZIs. Hence, various design methods [1], including matched filters, can be directly applied and programmed. (Such programming can also be simply remapped for use with (symmetric) binary tree or other self-configuring layer architectures [14]). Applications of such filters in telecommunications, for example, include dispersion compensation and gain equalization [1].

## 5. Non-redundant array single-layer filter

Many applications require narrow filter linewidths while also having moderately large free spectral range – for example, so that one spectral line can be distinguished or separated from a large number of others. Generally, narrow linewidths in optical spectrometers or filters require long path lengths or path length differences. High finesse resonators provide large effective path lengths because the light effectively makes many passes through the resonator, but tuning narrow resonances over large wavelength ranges can be challenging in integrated devices. Long path lengths are possible in waveguides, and long AWGs can show narrow effective linewidths [24]. However, physically defining the different waveguide lengths to high precision can be challenging, and large numbers of waveguides are required if the free-spectral range is to be a correspondingly large number of linewidths. Self-configuration of a simple filter interferometer mesh as discussed above can avoid the necessity of very precise fabrication of waveguide lengths. However, using a large number of waveguides so that we could still have a large free-spectral range would require a correspondingly large number of interferometers, increasing complexity and physical size.

We propose a different approach to narrow linewidth and large free-spectral range, one that requires only relatively small numbers of waveguides, and that is well-suited to our architecture. A key to this approach is that the waveguide lengths are no longer uniformly spaced. With careful choices of lengths, the filter can have both narrow linewidths and a broad free-spectral range; the filter peak can even be continuously tunable. The price to pay is only that the rejection of other wavelengths is not perfect.

One strategy for choosing the waveguide lengths, a so-called non-redundant array (NRA) [26–30], is already well known for spatial arrays, such as microwave antenna arrays for measuring arrival direction in radar signals [29], in spatial interferometric approaches in astronomy [28] or in optical phased arrays [27]. The idea of an NRA in spatial applications is that the spacing between any pair of elements (such as antennas) is different from the spacing between any other pair of elements. As a result, the interference pattern formed from any two elements has a different spatial frequency (or, in two dimensions, **k** vector) from that from any other pair. Hence, generally, these interference patterns tend mostly not to add constructively. However, we can choose the phase of all the antennas so that, in one chosen direction, the result is constructive. The result is a strong (angular) peak whose width tends to be given essentially by the largest antenna separation, even though we are using only a sparse array of actual antenna elements. The price for this is that there is some generated or detected signal at the other directions, though this may be much weaker, especially as the number of antennas (or, in our case, waveguides) is increased.

In one dimension, one approach for generating NRA separations is a so-called Golomb ruler [26,31–33]. An example Golomb ruler with four markings, at positions 0, 1, 4, and 6, has different separations between any different pairs of markings. For example, 0 and 1 are obviously separated by 1 unit, 4 and 6 by 2 units, 1 and 4 by 3 units, and so on. (In this, “perfect” Golomb ruler case, all integer separations from 1 to 6 exist.)

Here, we exploit this idea in the spectral domain for optical filters. (Incidentally, the first proposal of the NRA concept was in the spectral domain for avoiding intermodulation at radio frequencies [26].) Our waveguides now have lengths set by such an NRA principle, but the architecture is otherwise the same as in Fig. 1. This approach allows a strong and narrow spectral peak, albeit accompanied by some finite transmission at other wavelengths.

Fig. 3(c) shows example behaviors for a device with a set of 16 waveguides of relative lengths $m_p \delta l_o$ with the $m_p$ values for successive waveguides being the successive elements of the (Golomb ruler) set of integers $\{0,1,4,11,26,32,56,68,76,115,117,134,150,163,168,177\}$, where we use the same $\delta l_o = 16.51 \mu m$ as for Fig. 3(a). Fig. 3(d) shows the curve for $\gamma = \pi / 2$ at finer frequency resolution to show the character of the peak response more clearly.

The frequency when all waveguide outputs are added in phase by the mesh corresponds to the peak of the spectral response. In an ideal loss-less case, all the power at that frequency is coupled to the signal output waveguide, and this response can also be self-configured as before when shining in the frequency of interest. Though we have chosen explicitly non-redundant sets of integers to define waveguide lengths, one could also likely get similar response with other, and possibly non-integer, lengths provided those lengths spanned a similar set of waveguide lengths, possibly even in a somewhat random fashion.

We see that this approach would allow very narrow bandwidths and wide free-spectral range even with limited numbers of waveguides, at the expense of limited rejection of other wavelengths. It is well-suited to our approach that can compensate automatically for minor fabrication errors (especially in long waveguides) and allows simple tunability.

## 6. Multilayer filters

### *Unitary multilayer filters*

By adding more interferometer mesh layers, as in Fig. 2, we can program up to $N$ orthogonal filter functions to give corresponding outputs in the signal output of the corresponding layer. All filter functions, one from each layer, are available simultaneously, without any loss beyond the basic background losses of the system – that is, with no additional "splitting" loss. Even with unequally spaced waveguide lengths, as in the non-redundant array filter above, we can still losslessly and simultaneously form filter functions up to any $N$ orthogonal linear combinations of the corresponding Fourier-like basis functions $h_p(\delta\omega)$ as in Eq. (4).

Because of the construction of the multilayer unitary filter, the multiple filter functions it implements are physically guaranteed to be orthogonal to one another (at least if the mesh is "perfect" – that is, with 50:50 beam beamsplitters and the same loss on all paths through the mesh, a condition that can be arranged even with imperfect fabricated devices [13,25]); each such filter function corresponds to a different row of a unitary matrix, and such rows are necessarily orthogonal (see Supplementary section S1 for discussion of orthogonality). See Supplementary section S3 for an extended discussion and self-configuring and programming approaches.

For example, with an $N = 16$ waveguide design as in the single-layer simple filter above (Fig. 3(a)), but extended to a 15 layer mesh, we could have 16 filter responses, each like the filter response of, say, the $\gamma = 0$ design in Fig. 3(a), but shifted to be equally spaced throughout the spectral range of the filter. Each such filter peak would line up spectrally with the various zeros of every other filter response. This kind of filter would correspond to the usual outputs of a corresponding AWG filter. Note, though, that we could arbitrarily choose which such filter response appeared out of which output. So, in general, unlike an AWG, we could also perform arbitrary permutation of the channels among the outputs, allowing the device also to function as a switch. Quite generally, any set of up to $N$ orthogonal filters formed from the Fourier-like basis set $h_p(\delta\omega)$ as in Eq. (4) is possible, opening a wide range of simultaneous filter functions.

### *Rejection filter*

One interesting multilayer filter would be a rejection filter designed to eliminate one or more wavelengths from a spectrum; for example, one might want to eliminate a laser line in Raman spectroscopy. We could, for example, self-configure the first row of the mesh by shining in the frequency to be eliminated. In an ideal mesh, all that light would then be coupled out of the first signal output of the mesh. Non-idealities would limit the rejection possible this way, but the use of "perfect" MZIs [13,25,34] and related techniques are promising for high rejection ratios (e.g., > 80dB in first experiments in meshes for spatial applications [34]). The remaining light would then pass into subsequent layers for further spectral analysis. Of course, that first layer, because its spectral response cannot be infinitely sharp, would also remove some light at other frequencies, which would somewhat distort the remaining spectra. The use of a non-redundant array, as in Fig. 3(b), could allow spectrally sharp filtering of light to be rejected, at the cost of some loss and a significant and "rough" nonuniformity in the remaining spectral response.

One could use the multiple layers in the mesh to filter out multiple different frequencies. With the first layer configured to remove essentially all of some frequency $f_1$, incident light at some other frequency $f_2$ would then presumably mostly pass through into the second layer. That second layer could then be configured to remove essentially all the remaining power at frequency $f_2$. Note that these two layers, taken together, can remove all the power of any two such frequencies. (They do not necessarily separate all of frequency $f_2$ to the second output – some of that frequency may also emerge from the output of the first layer – but all the power at both frequencies will be removed.)

We could proceed similarly through other layers of the mesh. A full mesh, with $N-1$ layers and $N$ input waveguides could remove any $N-1$ frequencies from the spectrum; the remaining spectrum would emerge from the $N^{\text{th}}$ waveguide. Note this technique can be used for any $N-1$ frequencies in the free-spectral range, at the cost of some spectral non-uniformity and loss in the remaining transmitted light in the $N^{\text{th}}$ waveguide.

Such programmable filters could be used to block or separate out specific wavelengths. For example, in astronomy, we may want to block light-polluting sodium or mercury lines from street lights, and block or separate various specific atomic lines, such as hydrogen alpha or beta. Importantly, the filter could be tuned for different such rejections or separations – the filter function need not be fixed. Note the light filtered this way is still available for analysis; it is not absorbed but rather separated out. By calibrating the device, specific spectral line positions or red shifts could also be measured.

### *Non-unitary multilayer filter*

We can implement fully programmable non-unitary matrices $M$ – for example, using the singular-value decomposition (SVD) architecture [9] of two unitary meshes with a line of modulators between them (see Supplementary section S3 for an extended discussion). Then we can implement multiple non-orthogonal filtering functions simultaneously on the same incoming light.

Such an architecture obviously cannot violate basic laws of physics. Unless we incorporate gain in the mesh (for example, adding that to the modulation mechanism), we cannot, for example, have two different filter functions that transmit 100% of the input light at the same frequency to two different outputs at the same time. In matrix terms, without gain no singular value in the decomposition can exceed unity in magnitude. We can, however, simply rescale any desired matrix by dividing it by its largest singular value; in that rescaled case, arbitrary sets of non-orthogonal filter functions are possible, at least to the extent they can be represented on the Fourier-like basis as in Eq. (3). We should, however, expect some overall loss in such non-orthogonal filters – power is formally "dumped" in the modulators that represent singular values of magnitude less than unity. See Supplementary section S3 for an extended discussion.

## 7. Measurement of the temporal coherency matrix

Light that contains a range of frequencies – such as continuous spectra from a thermal light source, a light-emitting diode, or some kinds of laser pulses – can usefully be described as (temporally) partially coherent. Such partial coherence is well understood theoretically (e.g., Ref. [35]). The mode amplitude $x(t)$ in a single-mode waveguide, such as the input waveguide in Fig. 1(a), can then be described in terms of its (temporal) mutual coherence function $\Gamma(t_1, t_2)$ for any two times $t_1$ and $t_2$. For stationary processes, this function depends only on the time difference $\tau = t_1 - t_2$, and for ergodic processes, what would formally be an ensemble average of over a statistical ensemble of possible functions $x(t)$ can instead be evaluated as a time average. Such stationarity and ergodicity are typically assumed for partially coherent light [35]. Then, formally,

$$\Gamma(\tau) = \left\langle x^*(t) x(t+\tau) \right\rangle_t \tag{8}$$

where $\langle \cdot \rangle_t$ denotes averaging over time *t*. With relative waveguide time delays that can be written as $p\delta t_o$ for waveguide *p* in an AWG-like waveguide array of *N* waveguides, we then have access to $\tau$ values of the form $(s-p)\delta t_o$ for the relative time delay between waveguide *s* and waveguide *p*. So, we can also write $\Gamma$ in the form of a "coherency matrix" with matrix elements

$$\Gamma_{ps} \equiv \Gamma\left((s-p)\delta t_o\right) \tag{9}$$

If we can establish these matrix elements for the field, then we have evaluated the (temporal) mutual coherence function, at least at the discrete set of points in time delay allowed as the integers *s* and *p* range from 1 to *N*.

Just as for other filter functions, the waveguide array has essentially mapped from a vector of amplitudes separated in time to a vector of amplitudes in space, in the different mesh input waveguides. Some of us previously showed how to measure the (spatial) coherency matrix of a light field using a multiple-layer self-configuring mesh [19] as in Fig. 2, sequentially optimizing the output powers in successive layers. We can now take a similar approach to measuring this temporal coherency matrix and hence the mutual coherence function. Effectively, we map temporal coherence, which can loosely be described as the degree to which light amplitudes at different times could interfere with one another, to spatial coherence, which equivalently asks how light amplitudes at different points in space or in different waveguides could interfere with one another.

We repeat the essence of the previous approach [19]. We first adjust the elements in the first mesh layer to maximize the power in its signal output. This will have established and measured the first eigenfunction of the coherency matrix; the signal output power will the first (largest) eigenvalue, and the settings of the interferometers in the layer give the corresponding eigenvector. We proceed similarly through the successive mesh layers, establishing all the eigenvectors and eigenvalues in decreasing order. By this means, we have effectively measured the entire coherency matrix as in Eq. (9). (In addition, the use of a non-redundant array of waveguide lengths would allow us to measure the mutual coherence for a larger number of different time delays with the same number of physical measurements.) We have also physically separated the partially coherent field into its mutually incoherent and orthogonal components, presenting these as the signal output powers. (This formally corresponds to separating the field into its Karhunen-Loève expansion.) We are not aware of another approach that both measures the mutual coherence and separates it physically into these components. This separation is also non-destructive; once set up, all the subsequent light passes through from the input to the outputs. A detailed derivation of this approach and the non-redundant extension can be found in Supplementary Section S4.

## 8. Other filter modalities and extensions

Because of the flexible programmability of this filter approach and the further possibilities allowed by self-configuration, there are many additional possible operating modes. There are also some extensions we can make to the physical architecture. Here we mention some of these briefly, with an extended discussion in Supplementary section S5.

(a) The device can be run backwards for (i) wavelength multiplexing, (ii) as a tunable mirror – e.g., as a laser tuner, or (iii) as a pulse shaper or generator.

(b) When operated for separating partially coherent fields, the device can (i) automatically separate high-coherence sources (such as light from different lasers), sharp spectral lines, or different wavelength channels, without prior knowledge of their wavelengths, and (ii) can look for and measure absorption lines or spectra against uniform background light.

(c) The architecture of the device can be extended or modified in several ways. (i) Though we have mostly discussed either single layer devices or complete unitary meshes with $N$ mesh input waveguides, the device can operate with fewer – say, $Q$ – layers, allowing a less complicated photonic circuit while still usefully separating $Q$ wavelengths at high resolution, for example. (ii) Though we have discussed self-configuration based on power optimization, optimization on other measurable parameters is possible, with one interesting example being to optimize based on "eye-opening" or minimum bit-error rate in telecommunication systems, which could effectively construct or self-configure optimum matched filters. (iii) Though we show a fixed power splitter in Fig. 1 and Fig. 2, a controllable power splitter could be substituted, which could be useful when operating with waveguides of very different lengths (and hence different losses). (iv) Components in the meshes will not generally be perfect; such imperfections can be compensated and/or rejection improved by using "perfect" MZI approaches or the use of multiple layers for a given rejection [13,25,34]. (v) By dithering waveguide phase shifts with small modulations, the device can perform derivative spectra, which is often a useful mode for weak signals or for rejecting background. Finally, (vi) by using multiple input waveguides, each with its own waveguide array, all feeding into one or more larger mesh(es), combined spatial and spectral operation is possible simultaneously, as might be interesting for compensating modal dispersion, for example.

## 9. Discussion

In a fair comparison to conventional AWG filters, when our approach is configured as a simple filter, as in the results of Fig. 3(a), the filtering performance is essentially the same as that of an AWG with similar waveguides. Also, some previous AWG approaches [8] have already proposed some degree of programmability and tunability based on a set of phase shifters in the waveguides. Furthermore, our approach requires significant added complexity in the many MZIs in the photonic circuit and requires corresponding electronic circuitry to drive the MZI phase shifters. Because of that additional complexity, the use of large numbers of AWG waveguides, as used in high-performance AWG circuits, could be technologically challenging or limiting in our approach.

We believe, however, that our approach offers many potential benefits for filtering and spectroscopy more broadly, justifying the additional complexity beyond conventional AWGs. The benefits of our approach lie in several areas.

First, it offers a broad range of different filter functions. Even the simple single-layer filter can implement any finite-impulse response filter (e.g., transversal, tapped-delay-line, or matched filters) that is based on the set of time-delays of the waveguides. (Such flexibility contrasts with resonator filters, which tend to have fixed and limited Q-factors and linewidths.) Design can be as simple as Fourier decomposition of the desired filter function. Multilayer meshes allow multiple simultaneous filtering functions, each of which can be arbitrarily set across the range of possible filters, with no additional fundamental or "splitting" loss introduced for simultaneous orthogonal filters. (Only non-orthogonal filters necessarily require some additional loss.) All the filters are fully tunable over an entire free-spectral range, requiring only the same phase shifters used for the basic programmability of the mesh.

Second, filter functions can be automatically discovered by self-configuring to the input light. For a single input wavelength, simple progressive algorithms can set a mesh layer to filter exactly that wavelength, without any prior knowledge of the wavelength. By leaving the configuration algorithm running, the filter can automatically track that wavelength if it changes for any reason. Equivalently, periodically recalibrating with a reference wavelength can hold or reset the device behavior even in the presence of environmental changes (e.g., temperature) or component aging. Such a single wavelength input also allows simple calibration of the set of waveguides and the entire set of mesh layers by completely progressive algorithms based only on power minimization or maximization in a detector or detectors [14,36]. Such calibration also allows automatic compensation for fabrication errors in the precise lengths of waveguides and any other static phase shifts, allowing possibly very long waveguide lengths for precise spectral filtering. Variations in loss in different waveguide paths can also be automatically compensated by the mesh layers. Pre-compensation with a programmable power splitter allows simple filter analysis and programming even with substantially different losses in different waveguides. (In this case, the spectral resolution of the filter is not limited by losses in the waveguides.)

Third, this can be a universal filter approach that is software-defined and is completely programmable and reprogrammable in the field with simple progressive algorithms.

Fourth, this approach offers novel operation modalities. In an $N$ waveguide device, up to $N-1$ wavelengths lying anywhere in the filter's free-spectral range can be completely and automatically rejected from the remaining output. Our additional novel concept of a non-redundant array filter allows very narrow and programmable spectral response even with only moderate numbers of waveguides and circuit complexity, at the expense only of finite rejection at other wavelengths. The use of multilayer meshes together with algorithms that maximize (or minimize) power over an entire mesh layer at once allows powerful functions operating on partially coherent light, including separation into its mutually temporally incoherent components (a physical Karhunen-Loève decomposition). This approach allows measurement of the temporal coherence function; we are not aware of any other approach that accomplishes either this physical separation or this non-destructive measurement of temporal coherence. It also enables functionalities such as absorption spectroscopy in the presence of a broad background spectrum.

## 10. Conclusions

In conclusion, the approach to spectral filtering presented here, by exploiting the programmability of interferometer meshes together with waveguide arrays, offers a wide range of filters and operational modalities. Multiple simultaneous filters are possible. All these filter functions can be programmed and re-programmed after fabrication and can exploit the self-configuring possibilities of forward-only meshes formed from self-configuring layers to set themselves up optimally based on the input light. Some of the resulting capabilities, such as measurement and separation of partially coherent light fields, are apparently beyond previous optical systems. The approach also offers tolerance to fabrication variations, self-calibration, and self-stabilization. Taken together, these features and capabilities are very promising for future integrated photonic filters and spectrometers.

## Funding

DM and SF acknowledge funding from the Air Force Office of Scientific Research (AFOSR) grant FA9550-21-1-0312), and DM also from AFOSR grant FA9550-23-1-0307. CR-C is supported by a Stanford Science Fellowship. CV acknowledges support from the Texas Instruments Stanford Graduate Fellowship. AK acknowledges support from the John and Kate Wakerly Stanford Graduate Fellowship and the NSF Graduate Research Fellowship Program. MV acknowledges funding from the European Union Horizon 2021 Programme under the Marie Skłodowska-Curie Actions (agreement no. 101067268). OS acknowledges support from NASA through the NASA STMD Early Career Initiative (ECI) program (investigator – Dan Sirbu).

arXiv:2501.11811

# Universal programmable and self-configuring optical filter: Supplemental document

David A. B. Miller*[,1], Charles Roques-Carmes[1], Carson G. Valdez[1], Anne R. Kroo[1], Marek Vlk[1,2], Shanhui Fan[1] and Olav Solgaard[1]

[1]*Stanford University, Ginzton Laboratory, 348 Via Pueblo Mall, Stanford CA 94305*
[2]*Department of Physics and Technology, UiT The Arctic University of Norway, NO-9037 Tromsø, Norway*
**dabm@stanford.edu*

We give extended mathematical analysis and discussion of the approach to filtering discussed in the main text and an expanded discussion of some extensions of device usage and architectures. S1 gives a detailed formalism for analyzing the filter behaviors. S2 gives a complete analysis of a simple single-layer filter. S3 gives extended discussion of the classes and range of possible filter functions, especially for multilayer filters, and of approaches for programming the necessary matrices onto the interferometer mesh. S4 gives a detailed discussion of the analysis of temporal coherence and its measurement by the system. S5 discusses some other operating modes and extensions in more detail.

## S1. Formalism for analyzing filter behavior

To understand this filter approach in detail, we can set up some physical and mathematical formalism. Much of this is standard for analyzing devices such as arrayed waveguide grating (AWG) routers, but we give this in relatively complete form because we need to extend it, especially for waveguide lengths beyond those used in conventional AWGs and for analysis of multiple-layer interferometer meshes, which allow many novel filters.

### General description

Consider the system as illustrated schematically in Fig. S1, which is a generalized version of Fig. 1(a) of the main text.

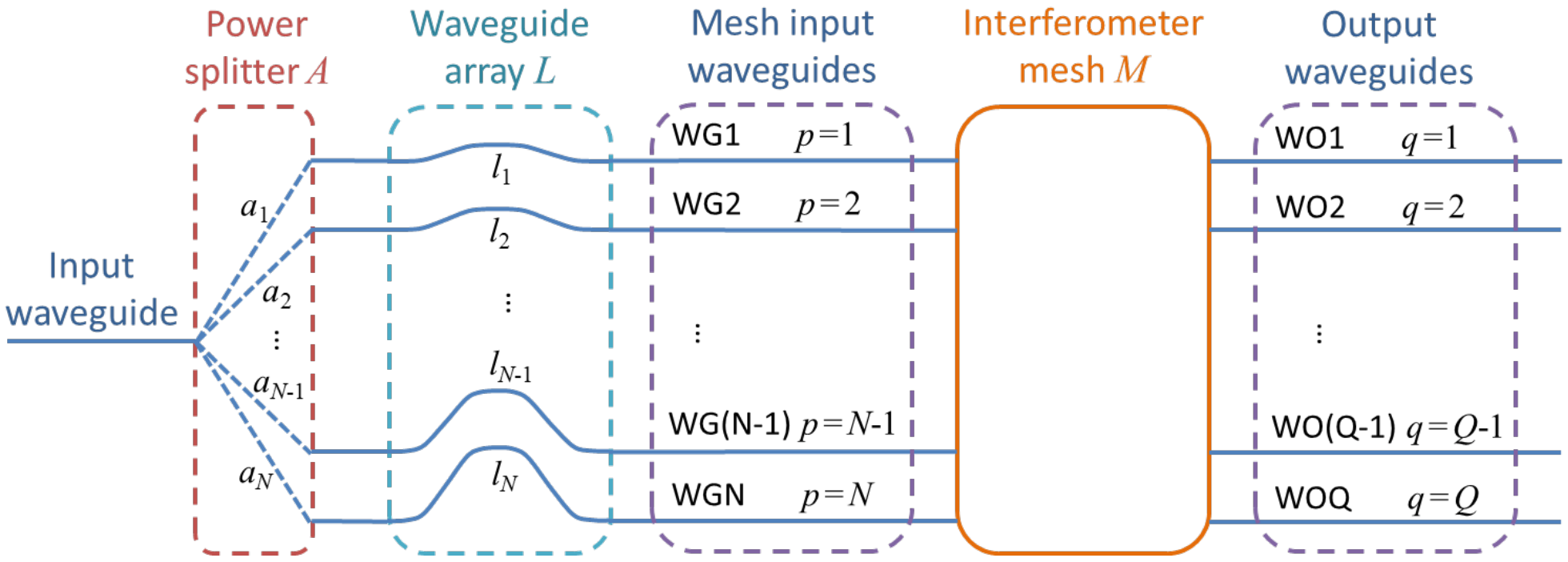

Fig. S1. General block diagram of the device.

All waveguides are presumed to be single-spatial-mode guides. We presume a single input waveguide. In this analysis, we consider one wavelength or frequency at a time, so, some angular frequency $\omega$, which would correspond to a conventional frequency $f = \omega / 2\pi$, and a free-space wavelength $\lambda = c / f = 2\pi c / \omega$ where $c$ is the velocity of light in free space. Since this is a linear, time-invariant system (at least over the time-scales relevant for optical frequencies), different frequencies are never mixed to generate new frequencies within the system, so the response of the system to a sum of light at different frequencies is just the sum of the responses to the individual frequencies.

## Power splitting

An input field at this angular frequency $\omega$ and of amplitude $x$ in the single input waveguide is split by this power splitter $A$ among the waveguides in the array $L$ with amplitude coefficients $a_1$ to $a_N$. In general, this split need not be equal, but if it were, for a lossless splitter these amplitudes would all be

$$a_p = 1/\sqrt{N} \tag{S1}$$

in magnitude; then the relative powers, presumed proportional to the modulus squared of the amplitudes, would add to 1. Whether or not the split is equal, quite generally, for a lossless system by power conservation we write

$$\sum_{p=1}^{N} \left|a_p\right|^2 = 1 \tag{S2}$$

We have written the modulus squared here (i.e., $|a_p|^2$ ) in each case, though in fact, without loss of generality, we could take these coefficients $a_p$ to be real because we will have other points at which we can add arbitrary phases mathematically if we need to. (It will not matter in the final operation if there are different fixed phases associated with these splitting amplitudes; such phases can readily be calibrated out of the system [1].) If there is some common loss in the waveguides or the power splitter, we can simply presume some constant loss factor for the entire system and proceed otherwise as if the system is lossless.

## Waveguide lengths

The waveguides in the array $L$ will be fabricated to a set of lengths $l_1$ to $l_N$, which we could take as the waveguide lengths from the output of the power splitter to the input to the mesh. In the analysis that follows, the total length of the guides is of little or no interest; what matters is the difference in the lengths of the guides. (Lengths that are common to all paths just correspond to the same phase factor over all paths for a given frequency, and such a common phase factor is of no particular importance here.) For convenience, we can define such differences relative to some "center" length $l_c$. This could be the length of some "center" waveguide (if there is one), or some average length; in practice, we can just choose it to be some length that is convenient for our algebra. (It can also be the length of the first waveguide if that is more convenient for analysis.) Then we work with the relative length difference of waveguide $p$ compared to $l_c$. So, we use the quantities

$$\delta l_p = l_p - l_c \tag{S3}$$

Various parameters in the analysis, such as power-splitting coefficients, beamsplitter ratios, and controlled phase shifts, could all depend on frequency, in which case we would regard these parameter values as being the ones at the angular frequency $\omega$ of interest. As a first approximation and for simplicity of analysis, we presume these parameters do not depend on frequency or wavelength, at least for narrow bandwidths. We do retain frequency dependence of the effective refractive index of the waveguide mode, however, since this is important in the system behavior through group velocity effects, and we discuss this next.

## Waveguide phase delays and effective refractive indexes

In waveguides, some care is required with appropriate refractive index values, both for the effective refractive index at some frequency and with the effects of frequency-dependence or dispersion of that index. Especially in single-mode waveguides with small cross-sections, first, the effective (phase) refractive index for light in the guide at a given frequency, $n_r$, can differ significantly from the underlying refractive index $n_o$ of the core material itself [2], in part because some of the field amplitude is in the lower-refractive index waveguide cladding material. Second, the group index $n_g$ may be quite different again, especially because of strong modal dispersion effects in small guides.

To understand refractive index effects and phase delays in single-mode guides, first we can formally define the phase refractive index for the waveguide mode

$$n_r = \frac{c\beta(\omega)}{\omega} \tag{S4}$$

where $\beta(\omega)$ is the (phase) propagation constant of the mode at (angular) frequency $\omega$. Now we choose some "center" or reference (angular) frequency $\omega_c$ (corresponding to a conventional frequency $f_c = \omega_c / 2\pi$). This need not be in the actual center of some frequency band of interest, though conveniently it can be. For future use, we can define the effective wavelength inside the waveguide at the center frequency as

$$\lambda_c = \frac{c}{n_r f_c} \equiv \frac{2\pi c}{n_r \omega_c} \tag{S5}$$

We can usefully approximate $n_r$ by a first-order expansion of its behavior with respect to frequency as

$$n_r(\delta\omega) = n_c + \delta\omega \frac{dn_r}{d\omega} \tag{S6}$$

where $n_c$ is the (phase) index at the center frequency, $dn_r / d\omega$ is the rate of change of this refractive index $n_r$ with respect to the angular frequency $\omega$, and

$$\delta\omega = \omega - \omega_c \tag{S7}$$

is the (angular) frequency separation from the center (angular) frequency.

The relative phase delay in propagating through waveguide $p$ can then be written

$$\begin{aligned} \delta\phi_p(\delta\omega) &= 2\pi n_r(\delta\omega)\frac{\delta l_p}{\lambda} = n_r(\delta\omega)\frac{\omega}{c}\delta l_p \\ &= \left(n_c + \delta\omega\frac{dn_r}{d\omega}\right)\frac{(\omega_c + \delta\omega)}{c}\delta l_p \end{aligned} \tag{S8}$$

Dropping terms in $(\delta\omega)^2$, consistent with taking the first-order expansion as in Eq. (S6), Eq. (S8) leads to

$$\delta\phi_p(\delta\omega) = \left(n_c\omega_c + n_g\delta\omega\right)\frac{\delta l_p}{c} \tag{S9}$$

where

$$n_g = n_c + \omega_c \frac{dn_r}{d\omega} \tag{S10}$$

which is typically referred to as the group index. (Note that, if $n_r$ did not depend on frequency – so we would then have $n_r = n_c$ – then Eq. (S9) would become the simple expression $\delta\phi_p(\delta\omega) = (\omega / c) n_r \delta l_p$ .)

In the design of AWGs, it will often be the case that the waveguide lengths are chosen so that, at a center frequency,

$$n_c \omega_c \delta l_p / c = 2u_p \pi \tag{S11}$$

for some integers $u_p$, which is equivalent to saying that the lengths of the guides (compared to the "center" waveguide length) are integer numbers $u_p$ of $\lambda_c$ [the wavelength in the waveguide at the center frequency, Eq. (S5)], i.e.,

$$\delta l_p = u_p \lambda_c \tag{S12}$$

Such an approach leads to simpler analysis, and we take this approach here. Then Eq. (S9) can be rewritten in the simpler form

$$\delta\phi_p(\delta\omega) = n_g \delta\omega \delta l_p / c \tag{S13}$$

where we have dropped the integer multiples of $2\pi$ as in Eq. (S11) from Eq. (S9) since they make no difference when in the argument of any sine or cosine trigonometric function. In that case, only the group index would then be involved in the frequency response of the system [3].

In practice, it will be useful to work with length increments between guides that are themselves a specific integer number $m_o$ of $\lambda_c$. Such a choice will mean that the free-spectral range of the system (essentially, the useful spectral range – see below for a detailed discussion) will be a correspondingly smaller fraction of $f_c$. (For example, the width of the telecommunication C-band is ~ 1/44 of its center frequency, which would suggest using $m_o = 44$ to get a filter of an appropriate free-spectral range.) So, we can choose some underlying length increment for design

$$\delta l_o = m_o \lambda_c = \frac{2m_o \pi}{n_c \omega_c} \tag{S14}$$

We can now usefully write

$$\delta l_p = m_p \delta l_o \tag{S15}$$

where $m_p m_o = u_p$.

We can now also usefully introduce the effective time delay $\delta t_o$ for the wave at frequency $\omega_c + \delta\omega$ passing through that same waveguide length increment $\delta l_o$; specifically,

$$\delta t_o = n_g \delta l_o / c \tag{S16}$$

in which case Eq. (S13) becomes

$$\delta\phi_p(\delta\omega) = m_p \delta\omega \delta t_o \tag{S17}$$

For example, and for our explicit calculations, as mentioned in the main text, we presume a typical silicon photonics waveguide design that is a silicon strip waveguide 500 nm wide by 220 nm tall, with silicon dioxide cladding. We use material refractive indices of 1.444 [4] for the silicon dioxide cladding and $n_o = 3.4757$ [5] for the silicon core, as appropriate for 1550 nm free-space wavelength. Using an eigenmode solver (EMOpt [6]) with these parameters, at a "center" wavelength of 1550 nm we deduce a (phase) refractive index of $n_c = 2.45932$ and a group index of $n_g = 4.05647$ for the TE polarized mode in such a guide. The low phase index is in part because of penetration of the waveguide mode into the lower index cladding material surrounding that silicon core. The group index is strongly influenced by modal dispersion. (In calculations by others for other thin silicon guides [7], the group index for TE polarized modes can be as high as $n_g \simeq 4.25$ near 1550 nm (free-space) wavelengths.) So, in our analysis below, as in other analyses of AWG routers (see, e.g., [3]), we will include such dispersive effects in the waveguides, at least to lowest order. The group index also varies slightly over the wavelength range of interest (such as the telecommunications C-band), but for simplicity in our calculations we use the above group index value for all wavelengths in our calculations.

### Modeling the combination of splitter, waveguide array and interferometer mesh

We use the convention here that a forward propagating wave at angular frequency $\omega$ has the form $\exp[i(\omega t - \beta(\omega) z)]$, where $t$ is time and we take $z$ to be the appropriate distance along the waveguide. (If necessary, we can take the real part of any result to give a real field or amplitude.) The interferometer mesh in Fig. S1 can be represented by some matrix $M$ with matrix elements $M_{qp}$. This mesh and matrix will have $N$ input waveguides and $Q$ output waveguides. Here $p$ indexes the mesh input waveguides WG1 to WGN and $q$ indexes the output waveguides WO1 to WOQ. $N$ and $Q$ can be equal, but they need not be; $Q$ can also be greater than or less than $N$. A simple filter need only have one signal output waveguide (so we can have $Q = 1$).

For an amplitude at (angular) frequency $\omega = \omega_c + \delta\omega$ (as defined in Eq. (S7)) of value $x$ in the input waveguide, by definition the output field in the waveguide of index $q$ is given by

$$y_q(\delta\omega) = \left[\sum_{p=1}^{N} a_p M_{qp} \exp\left(-i\, m_p \delta\omega \delta t_o\right)\right] x(\delta\omega) \tag{S18}$$

where we have used the phase delay from Eq. (S17) in each waveguide.

In the language of frequency filters [8], we can rewrite Eq. (S18) as (Eq. (2) of the main text)

$$y_q(\delta\omega) = H_q(\delta\omega) x(\delta\omega) \tag{S19}$$

where (Eq. (3) of the main text)

$$H_q(\delta\omega) = \sum_{p=1}^{N} a_p M_{qp} \exp\left(-i\, m_p \delta\omega \delta t_o\right) \tag{S20}$$

is the frequency response of the filter as seen in output waveguide $q$. For analysis of some simple systems, we may presume equal power splitting $a_p = 1/\sqrt{N}$ as in Eq. (S1). Then for simplicity of notation, we rewrite the matrix elements as

$$\tilde{M}_{qp} = a_q M_{qp} = \frac{M_{qp}}{\sqrt{N}} \tag{S21}$$

(this will allow the matrix $\tilde{M}$ to be an actual unitary matrix in practice), giving the simpler form

$$H_q(\delta\omega) = \sum_{p=1}^{N} \tilde{M}_{qp} \exp\left(-i\, m_p \delta\omega \delta t_o\right) \tag{S22}$$

For the case where the $m_p$ are successive integers, the series in Eq. (S20) or Eq. (S22) essentially corresponds to a Fourier series, and $N$ is then the "order" of the filter [8]. Altogether, the system is implementing $Q$ different filter frequency responses $H_q$, with a possibly different such frequency response seen in each different (signal) output waveguide $q$.

Note in Eq. (S18), or equivalently Eq. (S20), the response at frequency $\delta\omega$ (relative to the "center" frequency $\omega_c$ ) is the same as the response at frequency $\delta\omega + 2\pi n / \delta t_o$ for any integer $n$; because $m_p$ is an integer,

$$\begin{aligned}
&\exp\left(-im_p\left[\delta\omega + \frac{2\pi n}{\delta t_o}\right]\delta t_o\right) = \exp\left(-i\left[m_p\delta\omega\delta t_o - 2\pi m_p n\right]\right) \\
&= \exp\left(-im_p\delta\omega\delta t_o\right)\exp\left(-i2\pi m_p n\right) = \exp\left(-im_p\delta\omega\delta t_o\right)\times 1 = \exp\left(-im_p\delta\omega\delta t_o\right)
\end{aligned} \tag{S23}$$

So, each term in the sum is the same if we add $2\pi n / \delta t_o$, which makes the response periodic in frequency with a "free-spectral range" of

$$\omega_{FSR} = 2\pi / \delta t_o \tag{S24}$$

or, in conventional frequency terms

$$f_{FSR} = 1 / \delta t_o \tag{S25}$$

or, explicitly, using Eq. (S16)

$$f_{FSR} = c / n_g \delta l_o \tag{S26}$$

So, in understanding what filter functions we can generate, we note that only one such free-spectral range matters (with the response in every other free-spectral range being identical). So, we restrict our analysis to the "zeroth" free-spectral range, which we can choose to regard as being from (angular) frequency $\delta\omega = -\pi / \delta t_o$ to $\delta\omega = \pi / \delta t_o$.

## Filter functions, their representations, inner products, and Hilbert space

Much discussion of linear filter behavior is based on approaches that consider functions that may be continuous over all times and all frequencies or may be based on infinite sums of discrete and equally spaced frequencies or times. Approaches that work in such cases include Fourier transforms and z-transforms. Our case is somewhat different. First, we generally only have finite numbers of underlying physical functions we are considering, with the finite number being related to the number of waveguides in our arrays. Second, we may not even have uniform spacings of time delays between our different waveguides. As a result, it can be clearer to consider the filter functions of interest as being more generally in Hilbert spaces that can incorporate these restrictions. (Hilbert spaces are often implicit in the common approaches to analyzing and designing filters, but discussions of Fourier and z-transforms typically do not formally need to emphasize or point out this underlying mathematics.) Our use of Hilbert spaces enables us to be quite explicit about what functions we can consider (they are all the ones that can exist in our Hilbert spaces), retaining underlying mathematics of orthogonality and functional analysis without presuming that full Fourier or z-transforms must be used or be applicable.

With this background, to start, consider two possible filter functions formed as in Eq. (S20)

$$\begin{aligned}
H_a(\delta\omega) &= \sum_{p=1}^{N} \mu_{ap} \exp\left(-i\, m_p \delta\omega\delta t_o\right) \equiv \left|H_a\right\rangle \\
H_b(\delta\omega) &= \sum_{p=1}^{N} \mu_{bp} \exp\left(-i\, m_p \delta\omega\delta t_o\right) \equiv \left|H_b\right\rangle
\end{aligned} \tag{S27}$$

where we can consider that we are replacing the product of coefficients like $M_{qp}a_p$ with a single coefficient like $\mu_{ap}$ or $\mu_{bp}$ in writing these expansions. (We will clarify the meaning and use of Dirac notations like $\left|H\right\rangle$ below.) Generally, we can view such functions as existing mathematically in some "filter-function" function (or vector) space. With the formal mathematical definition of an inner product, which we can choose as

$$\langle H_a | H_b \rangle \equiv \frac{\delta t_o}{2\pi} \int_{-\pi/\delta t_o}^{\pi/\delta t_o} \left[ \sum_{p=1}^{N} \mu_{ap}^* \exp\left(i m_p \delta\omega \delta t_o\right) \right] \left[ \sum_{j=1}^{N} \mu_{bj} \exp\left(-i m_j \delta\omega \delta t_o\right) \right] d\left(\delta\omega\right) \quad \text{(S28)}$$

such a function or vector space essentially becomes a Hilbert space, mathematically.

Since quite generally

$$\int_{-\pi/\delta t_o}^{\pi/\delta t_o} \exp\left(i u \delta\omega \delta t_o\right) \exp\left(-i v \delta\omega \delta t_o\right) d\left(\delta\omega\right) = \frac{2\pi}{\delta t_o} \delta_{uv} \quad \text{(S29)}$$

for any integers $u$ and $v$ and with the Kronecker delta $\delta_{uv}$, then the inner product reduces to

$$\langle H_a | H_b \rangle = \sum_{p=1}^{N} \sum_{j=1}^{N} \mu_{ap}^* \delta_{pj} = \sum_{p=1}^{N} \mu_{ap}^* \mu_{bj} \quad \text{(S30)}$$

Equivalently, we can formally define the set of "Fourier-like" basis functions

$$h_p\left(\delta\omega\right) = \sqrt{\frac{\delta t_o}{2\pi}} \exp\left(-i m_p \delta\omega \delta t_o\right) \equiv |h_p\rangle \quad \text{(S31)}$$

each of which is also normalized over a free-spectral range (that is, the inner product of the functions with themselves is 1, i.e., $\langle h_p | h_p \rangle = 1$). We call this a "Fourier-like" basis rather than just a Fourier basis because, first, we only have a finite number $N$ of orthogonal basis functions, rather than having $p$ run from $-\infty$ to $+\infty$ as in a conventional Fourier series expansion, and, second, in the general case, the $m_p$ need not be successive integers; for example, they might be given by a non-redundant array set such as a Golomb ruler, as discussed in the main text.

Then, any filter functions in our filter-function Hilbert space can be represented on this basis, Eq. (S31), by the vectors

$$|H_a\rangle \equiv \begin{bmatrix} \mu_{a1} \\ \vdots \\ \mu_{aN} \end{bmatrix}, \quad |H_b\rangle \equiv \begin{bmatrix} \mu_{b1} \\ \vdots \\ \mu_{bN} \end{bmatrix} \quad \text{(S32)}$$

When using a Dirac notation like this in which $|\alpha\rangle$ represents a column vector, then $\langle\alpha|$ represents a corresponding row vector with complex conjugated elements; formally $\langle\alpha| \equiv (|\alpha\rangle)^\dagger$, where quite generally in Dirac notation the superscript "†" denotes the Hermitian adjoint or conjugate transpose of a matrix or vector. In this Dirac notation, we see the vector product $\langle H_a || H_b \rangle$ [conventionally shortened to the notation $\langle H_a | H_b \rangle$ as in Eq. (S30)] gives the same answer as the definition in Eq. (S28). Whether we use the representations of Eq. (S27) with inner product definition Eq. (S28) or of Eq. (S32) with inner product definition as the vector product $\langle H_a || H_b \rangle \equiv \langle H_a | H_b \rangle$, the inner product $\langle H_a | H_b \rangle$ lets us define that two non-zero frequency response functions $H_a$ and $H_b$ are orthogonal if and only if $\langle H_a | H_b \rangle = 0$.

With the orthogonal basis functions as in Eq. (S31), we can also rewrite the filter functions of Eq. (S27) in the form

$$\begin{aligned} H_a(\delta\omega) &\equiv \sqrt{\frac{2\pi}{\delta t_o}} \sum_{p=1}^{N} \mu_{ap} \left| h_p \right\rangle \equiv \left| H_a \right\rangle \\ H_b(\delta\omega) &\equiv \sqrt{\frac{2\pi}{\delta t_o}} \sum_{p=1}^{N} \mu_{bp} \left| h_p \right\rangle \equiv \left| H_b \right\rangle \end{aligned} \tag{S33}$$

Incidentally, a major reason for using Dirac notation, in addition to its algebraic simplicity and convenience, is that it allows us to refer to a function in the Hilbert function space in a way that is independent of what representation (basis set) we are using to specify the function. The representation in Eq. (S27) is in terms of the (infinite) set of amplitudes corresponding to the values of the continuous variable $\delta\omega$ within a free spectral range, whereas the representation in Eq. (S33) is in terms of an expansion over a discrete and countable basis set $\left| h_p \right\rangle$. Whichever way we represent it, we are referring to the same function, e.g., $\left| H_a \right\rangle$ or $\left| H_b \right\rangle$, in the mathematical Hilbert space of frequency response functions.

Because we can represent any frequency response function in the space of possible frequency response functions for this apparatus as a (weighted) sum over the $N$ orthogonal functions $h_p(\delta\omega) \equiv \left| h_p \right\rangle$, the Hilbert space of possible frequency response functions is $N$-dimensional. The Fourier-like basis as in Eq. (S31) is one possible convenient basis, but of course we can choose any other $N$-dimensional orthogonal basis formed by a unitary transform from this basis. With such basis sets we can represent any filter function that can be formed from this set of $N$ waveguides with their specified lengths $\delta l_p$ as in Eq. (S15). One other such useful basis could be the "sinc-like" functions we would obtain as the frequency responses of AWG outputs (see the discussion below for the analysis of a simple filter. Another other particularly interesting basis would be the set of orthogonal functions that are the eigenfunctions of the (temporal) coherency matrix, as we discuss below in relation to partially coherent light.

## Design of matrix elements to implement a filter function

If we know what filter function $\left| H \right\rangle$ we want, then we can find the best implementation of it, on the basis given by the functions as in Eq. (S31), by formally projecting it onto this basis to establish the coefficients $\mu_p$. Explicitly

$$\mu_p = \sqrt{\frac{\delta t_o}{2\pi}} \left\langle h_p \middle| H \right\rangle \equiv \sqrt{\frac{\delta t_o}{2\pi}} \int_{-\pi/\delta t_o}^{\pi/\delta t_o} \exp\left(i m_p \delta\omega \delta t_o\right) H(\delta\omega) \, d(\delta\omega) \tag{S34}$$

Note that, as mentioned above, the integers $m_p$ could be just consecutive integers for each successive integer $p$ (leading to successive waveguide lengths $\delta l_p$ that are be equally spaced, as in a conventional AWG), but they need not be. In the case of the non-redundant array filters in the main text, these $m_p$ can be quite widely and relatively arbitrarily spaced integers. Nonetheless, the process described here, with the coefficients as in Eq. (S34) still gives us the best possible filter design (in a "least squares" sense) on the resulting basis. In general, linear combinations of basis functions of the form in Eq. (S31) tell us what filter functions are possible and hence, formally, in what filter-function Hilbert space we are working.

## Time-domain description

Though we will mostly use the frequency-domain description as discussed above, we can briefly note the time-domain description. We use a time-domain discussion below when considering the (temporal) mutual coherence function and (temporal) coherency matrix.

It is straightforward to write the impulse response of the filter directly. If we presume a (Dirac) $\delta$-function input at time $t = 0$, then the output of one row $q$ of the filter in time is simply

$$\tilde{H}_q(t) = \sum_{p=1}^{N} a_p M_{qp}\, \delta\left(t - m_p \delta t_o\right) \tag{S35}$$

(Here we are neglecting some overall propagation time $t_{prop}$ through the system for mathematical simplicity, and because it likely will not matter to us in applications; we could incorporate that by subtracting it inside the argument of each $\delta$-function, delaying the output accordingly.) All we are saying here is that, if the input to the device is a short impulse (written mathematically as a $\delta$-function), then the output is a string of $N$ $\delta$-functions, delayed by times $m_p \delta t_o$, and with corresponding amplitudes $a_p M_{qp}$. This time-domain response, Eq. (S35), with appropriate normalization, is the Fourier transform of the frequency-domain response in Eq. (S20) (Eq. (3) of the main text).

### S2. Analysis of a simple filter

One filter construction, common in AWGs [9–14], is where the power is split equally between the waveguides as in Eq. (S1) and where the successive waveguides increase in length by equal amounts $\delta l_o$. For the moment, we are interested in only one "row" $q$ of the matrix, which is equivalent to considering only one (signal) waveguide output at a time; for example, with just a single "layer" of MZIs in the device, as in Fig. 1(b) or (c) of the main text, the matrix would anyway have only one row (so, obviously, then $q = 1$). We call this a simple filter.

We will formally "center" these length differences to get an average guide length difference of zero for mathematical convenience. (The guides themselves will, of course, all have positive lengths; we are essentially just subtracting off the length of the "center" or average guide.) So, we write

$$\delta l_p = \left[ p - 1 - \frac{(N-1)}{2} \right] \delta l_o \equiv \left[ p - \frac{(N+1)}{2} \right] \delta l_o \tag{S36}$$

(Note that formally, for even $N$, this would mean that $\delta l_p$ could be a half integer number of the length increments $\delta l_o$, rather than an integer number as formally presumed in Eq. (S15). However, the lengths $\delta l_p$ will still be spaced by integer numbers of length increments $\delta l_o$ since $p$ is an integer; so, this possible additional $\delta l_o / 2$ just corresponds to an overall phase factor common to all elements in the sums over the waveguide responses, and such overall phase factors are generally of no particular significance in the operation of the device. This does, however, lead to a minor complicating detail in discussing free-spectral ranges below.)

From Eq. (S17), therefore, we have relative phase delays in the guides of

$$\delta\phi_p(\delta\omega) = \left( p - \frac{(N+1)}{2} \right) \delta\omega \delta t_o$$

As mentioned in the main text, we can usefully add a controllable "tilt" to the input phase shifters of the mesh (e.g., PS1 to PS4 in Fig. 1 of the main text), adding a phase increment $\gamma$ between each successive waveguide, where $\gamma$ is a real constant that can be chosen to be positive, negative or zero. This will allow us to tune the "center" frequency of the filter after we set it up. Here, for algebraic convenience (but with no change in the underlying physics), we center the phase tilts around the middle of the set of input waveguides. So, instead of the form $\delta_p = m_p \gamma$ for this "tilted" phase delay as in Eq. (7) of the main text, with our simple choice of $m_p = p$ and this centering, instead we use a "tilted" phase delay

$$\delta_p = \left( p - \frac{(N+1)}{2} \right) \gamma$$

We choose uniform matrix element magnitudes for this filter. With our presumed equal power splitting, and using the simplified notation of Eq. (S22) where we incorporate the corresponding $\sqrt{1/N}$ splitting amplitude into the matrix elements, those elements in our single row $q$ of interest are then

$$\tilde{M}_{qp} = \frac{1}{\sqrt{N}} \exp\left( -i \left( p - \frac{(N+1)}{2} \right) \gamma \right) \tag{S37}$$

Substituting these matrix elements in Eq. (S22) gives

$$\begin{aligned} y_q(\delta\omega) &= \frac{1}{N} \left[ \sum_{p=1}^{N} \exp\left( -i \left[ p - \frac{(N+1)}{2} \right] \gamma \right) \exp\left( -i \left( p - \frac{(N+1)}{2} \right) \delta\omega\delta t_o \right) \right] x(\delta\omega) \\ &= \frac{1}{N} \left[ \sum_{p=1}^{N} \exp\left( -i \left[ p - \frac{(N+1)}{2} \right] \alpha \right) \right] x(\delta\omega) \\ &= \frac{1}{N} \left[ \sum_{s=0}^{N-1} \exp\left( -i \left[ s - \frac{(N-1)}{2} \right] \alpha \right) \right] x(\delta\omega) \\ &= \frac{1}{N} \exp\left( i \frac{(N-1)}{2} \alpha \right) \left[ \sum_{s=0}^{N-1} \exp(-is\alpha) \right] x(\delta\omega) \end{aligned} \tag{S38}$$

where $s = p - 1$ and for convenience we have defined $\alpha = \gamma + \delta\omega\delta t_o$.

Now, the summation in Eq. (S38) is a geometric series of the form

$$\sum_{s=0}^{N-1} z^s = \begin{cases} \dfrac{1 - z^N}{1 - z}, \text{ for } z \neq 1 \\ N, \text{ for } z = 1 \end{cases} \tag{S39}$$

where $z \equiv \exp(-i\alpha)$. Hence, for $\alpha \neq 2m\pi$, where $m$ is any positive or negative integer or zero

$$\begin{aligned} y_q(\delta\omega) &= \frac{1}{N} \exp\left( i \frac{(N-1)}{2} \alpha \right) \left[ \frac{1 - \exp(-iN\alpha)}{1 - \exp(-i\alpha)} \right] x(\delta\omega) \\ &= \frac{1}{N} \frac{\exp(iN\alpha/2)}{\exp(i\alpha/2)} \left[ \frac{1 - \exp(-iN\alpha)}{1 - \exp(-i\alpha)} \right] x(\delta\omega) \\ &= \frac{1}{N} \left[ \frac{\exp(iN\alpha/2) - \exp(-iN\alpha/2)}{\exp(i\alpha/2) - \exp(-i\alpha/2)} \right] x(\delta\omega) \\ &= \frac{1}{N} \left[ \frac{2i\sin(N\alpha/2)}{2i\sin(\alpha/2)} \right] x(\delta\omega) = \frac{1}{N} \frac{\sin(N\alpha/2)}{\sin(\alpha/2)} x(\delta\omega) \end{aligned} \tag{S40}$$

We can also rewrite this result as

$$y_q(\delta\omega) = \frac{1}{N}\frac{\alpha/2}{\alpha/2}\left[\frac{\sin(N\alpha/2)}{\sin(\alpha/2)}\right]x(\delta\omega) = \frac{\alpha/2}{\sin(\alpha/2)}\frac{\sin(N\alpha/2)}{N\alpha/2}x(\delta\omega)$$
$$= \frac{\text{sinc}(N\alpha/2)}{\text{sinc}(\alpha/2)}x(\delta\omega) \tag{S41}$$

where we take the definition

$$\text{sinc}\, x \equiv \begin{cases} \frac{\sin x}{x}, & x \neq 0 \\ 1, & x = 0 \end{cases} \tag{S42}$$

or explicitly now for $\alpha \neq 2m\pi$

$$y_q(\delta\omega) = \frac{1}{N}\frac{\sin\left(N\left[\delta\phi_c + \gamma + \delta\omega\delta t_o\right]/2\right)}{\sin\left(\left[\delta\phi_c + \gamma + \delta\omega\delta t_o\right]/2\right)}x(\delta\omega) \tag{S43}$$

or equivalently

$$y_q(\delta\omega) = \frac{\text{sinc}(N[\delta\phi_c + \gamma + \delta\omega\delta t_o]/2)}{\text{sinc}([\delta\phi_c + \gamma + \delta\omega\delta t_o]/2)}x(\delta\omega) \tag{S44}$$

For $\alpha = 2m\pi$ , since

$$\exp\left(i\frac{(N-1)2m\pi}{2}\right) = \exp\left(i(N-1)m\pi\right) = (-1)^{m(N-1)} \tag{S45}$$

then

$$y_q(\delta\omega) = (-1)^{m(N-1)}x(\delta\omega) \tag{S46}$$

The periodicity of the behavior of this filter in frequency space is easier to see from the “sine” form, Eq.(S43). If we change the frequency of the input from $\omega_a = \omega_c + \delta\omega_a$ to $\omega_b = \omega_c + \delta\omega_b$ where $\delta\omega_b = \delta\omega_a + 4m\pi/\delta t_o$ for any integer *m*, then no matter what is the integer value of *N*, the relation between the input $x(\delta\omega)$ and the output $y_q(\delta\omega)$ is unchanged because we have simply added an multiple of $2\pi$ to the argument of each sine function ( $2Nm\pi$ on the sine in the numerator, and $2m\pi$ on the sine in the denominator). That would give a periodicity in angular frequency with a period or frequency increment $4\pi/\delta t_o$. While this is strictly correct, more commonly in filter response we may be interested in the periodicity of the “power” response of the filter, which would depend on the modulus squared of the output of the filter. In that case, we have a relation of the form

$$\left|y_q(\delta\omega)\right|^2 = \frac{1}{N^2}\frac{\sin^2\left(N\left[\delta\phi_c + \gamma + \delta\omega\delta t_o\right]/2\right)}{\sin^2\left(\left[\delta\phi_c + \gamma + \delta\omega\delta t_o\right]/2\right)}\left|x(\delta\omega)\right|^2$$
$$= \frac{1}{N^2}\frac{1-\cos\left(N\left[\delta\phi_c + \gamma + \delta\omega\delta t_o\right]\right)}{1-\cos\left(\delta\phi_c + \gamma + \delta\omega\delta t_o\right)}\left|x(\delta\omega)\right|^2 \tag{S47}$$

which is periodic with an angular frequency increment $2\pi/\delta t_o$. The apparent difference between the field and power frequency periodicities is because of the possible sign change as evident by the expression Eq. (S46). If *N* is even, the sign of the output changes as we change angular frequency by $2\pi/\delta t_o$, but the intensity response does not care about such a sign change. (This minor point on the apparent doubling of the free-spectral range is algebraically because we formally "centered" the waveguide lengths about the average, and hence for even values of $N$ we have introduced a $\delta l_o / 2$ shift in the waveguide lengths compared to those of Eq. (S15).) With this understanding, we can formally state that the free-spectral range of this filter is as defined above in Eqs. (S24) - (S26).

Within the free-spectral range, in addition to one frequency at which the input and output magnitudes are the same, there are $N-1$ frequencies at which the response is identically zero. If we presume that the frequency $\omega_0 = \omega_c + \delta\omega_0$ is such that

$$\delta\phi_c + \gamma + \delta\omega_0 \delta t_o = 2m\pi \tag{S48}$$

i.e., for some integer $m$

$$\delta\omega_0 = \frac{2m\pi - \delta\phi_c - \gamma}{\delta t_o} \tag{S49}$$

this frequency corresponds to a maximum output from the filter (both sinc functions in Eq. (S44) are then 1). Then the set of frequencies

$$\omega_s = \omega_c + \delta\omega_0 + s\frac{2\pi}{N\delta t_o}, \; s = 1, 2, \ldots, N-1 \tag{S50}$$

will all correspond to zero output, so these frequencies, which are separated by

$$\Delta\omega = 2\pi / N\delta t_o \tag{S51}$$

or a conventional frequency separation

$$\Delta f = 1 / N\delta t_o \tag{S52}$$

are all perfectly rejected by the filter (at least for the output corresponding to this matrix row).

We can think of these (angular) frequencies $\omega_0, \omega_1, \ldots, \omega_{N-1}$ (or the corresponding conventional frequencies $f_0, f_1, \ldots, f_{N-1}$ obtained by dividing by $2\pi$) and their shifted replicas in the other free spectral ranges as the reference comb for the system. (See, e.g., [15] for discussion of comb sources.) Note that, if we choose $\gamma_q = 2m\pi - \delta\phi_c$ for some integer $m$, we are effectively able to "center" the response of the filter round the chosen frequency $\omega_c$ just by imposing this "tilt" on the phase shifts of the mesh. This completes the formal analysis of the simple filter in the main text, with frequency responses as in Fig. 3(a) of the main text.

## S3. Classes and range of filter operations

### *Meshes and matrices*

With interferometer meshes, we can usefully separate the architectures and operations into those that implement unitary matrices (at least within some constant, such as an overall loss of the system) and those that implement non-unitary matrices. We can make mesh architectures in which we add loss, with controllable absorbers or by deliberately "dumping" power to external waveguides, which would generally make the corresponding matrix $M$ represented by them be non-unitary. Mesh architectures more typically start by having no added internal loss or external "dump" ports, in which case they are represented by unitary matrices, typically written as $U$ or $V$ or their Hermitian adjoints (conjugate transposes) $U^\dagger$ or $V^\dagger$ (which mathematically are also necessarily unitary).

One particularly useful way to represent or construct any arbitrary non-unitary matrix is to use the singular-value decomposition (SVD) (see, e.g., [16]), in which we can write any matrix $M$ as the matrix product

$$M = VD_{diag}U^{\dagger} \tag{S53}$$

where $U$ and $V$ are unitary matrices and $D_{diag}$ is a diagonal matrix with a set of (generally complex) numbers $s_j$, known as the singular values, as the diagonal elements.

Equivalently, we can think of the matrix as mapping each of a set of orthogonal "input" vectors $|\psi_j\rangle$ to corresponding orthogonal "output" vectors $|\phi_j\rangle$ with coupling amplitudes $s_j$. The functions $|\psi_j\rangle$ are the eigenfunctions of the operator $M^{\dagger}M$, and the functions $|\phi_j\rangle$ are the eigenfunctions of the operator $MM^{\dagger}$, with positive real eigenvalues $|s_j|^2$, in both cases.

These sets are unique for any given matrix (at least within any arbitrary linear combinations of degenerate eigenvectors) – there is exactly one set of orthogonal "input" functions or vectors that maps, one by one, to exactly one set of corresponding "output" functions or vectors. These pairs of eigenfunctions, which can be called the "mode-converter basis sets" [16,17] for matrices representing optical systems, generally have fundamental physical properties and meaning in optical systems, including physical laws that apply only to them (see, e.g., [16,18]). The existence of these pairs of "input" and "output" orthogonal functions also automatically implies that any linear optical system has a set of orthogonal channels "through" it.

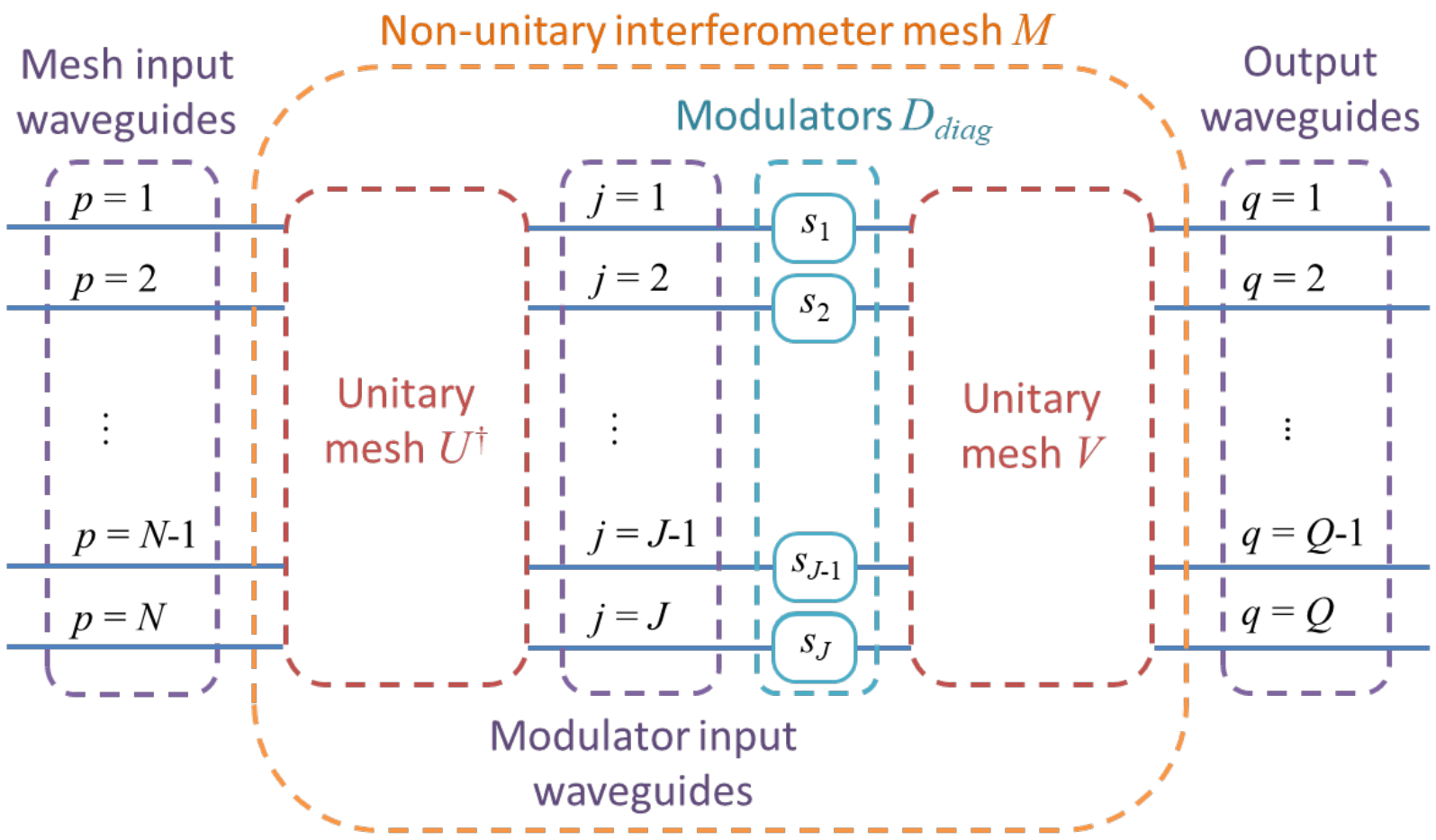

Fig. S2. Block diagram of a non-unitary matrix $M$ implemented by singular-value decomposition [19] onto a succession of a unitary matrix $U^{\dagger}$, implemented by a corresponding "input" unitary mesh, a diagonal matrix $D_{diag}$, implemented by a line of modulators, and a unitary matrix $V$, implemented by an "output" unitary mesh. Such a matrix $M$ need not be square; $Q$ can be equal to $N$, or can be greater than or less than $N$. The number of modulators $J$ need never exceed the smaller of $Q$ or $N$.

In this notation, the $|\psi_j\rangle$ are the corresponding columns of $U$ (and the corresponding $\langle\psi_j|$ are the rows of $U^{\dagger}$), and the $|\phi_j\rangle$ are the corresponding columns of $V$. In this notation, Eq. (S53) can be rewritten equivalently as

$$M = \sum_j s_j |\phi_j\rangle\langle\psi_j| \tag{S54}$$

Generally, then, any linear optical system between an "input" and an "output" space can be completely described by these coupled pairs of orthogonal functions $|\psi_j\rangle$ and $|\phi_j\rangle$ and their coupling strengths $s_j$, hence the term "mode-converter basis sets" [16,17]. If viewed as the set of communication channels between the input or "source" space and the output or "receiving" space, these same function pairs and coupling strengths can be viewed equivalently as the "communication modes" [16,20,21] of the system.

This SVD decomposition gives a general way to implement arbitrary matrices with interferometer meshes [19], as illustrated in Fig. S2. Each of the unitary matrices $U^\dagger$ and $V$ can be implemented with a unitary interferometer mesh and the matrix $D_{diag}$ can be implemented with a line of modulators between these two unitary meshes. The outputs of the $U^\dagger$ unitary mesh feed the inputs of the modulators, and the outputs of the modulators feed the inputs of the $V$ unitary mesh. The complex field transmission of a given modulator in waveguide $j$ is just chosen to be the singular value $s_j$.

The SVD architecture for implementing arbitrary matrices was introduced in Ref. [19], and has been employed by others subsequently (e.g., Ref. [22]). This method is only as complicated as it needs to be [23]; the number of adjustable parameters in the system overall (e.g., phase shifters) is essentially equal to the number of real numbers required to specify an arbitrary $Q \times N$ complex matrix.

### *Range of possible filter functions*

Given the ability to implement arbitrary unitary or non-unitary matrices with interferometer meshes, a wide range of filter functions can be implemented. Unlike some previous transversal or finite-impulse-response filters [24–26], we can contemplate a mesh with multiple outputs, each corresponding to a row of the matrix *M*, so the filter here can have multiple such frequency filtering behaviors simultaneously, one for each such mesh output. Unlike the fixed behavior of the outputs of a conventional AWG filter, we can have separate control over the filter functions for different outputs. This is an unusual capability for a physical filtering system, so we should clarify just what filters can be generated, both individually and in simultaneous sets.

As noted above, we can usefully view the filter frequency-response functions we can generate as formally being functions in a Hilbert space of possible functions. For *N* waveguides in the array, we can define this Hilbert space quite precisely as having the *N* dimensional basis set as in Eq. (S31) and the inner product as in Eq. (S28) or, more compactly, Eq. (S30), with frequency response functions that can be written as in Eq. (S27) or more compactly Eq. (S33).

In the discussion that follows, just as we did for the analysis of the simple filter above, for simplicity we presume equal power splitting, so all of the $a_p = 1/\sqrt{N}$ as in Eq. (S1), and we will fold these $a_p$ into the rescaled matrix elements $\tilde{M}_{qp}$ for simplicity of notation, as in Eq. (S22). We will also presume waveguide length differences based on integer multiples of an underlying length $\delta l_o$, which will give all such filters a free-spectral range as in Eq. (S51), $\Delta\omega = 2\pi / N\delta t_o$, or equivalently as in Eq. (S52), $\Delta f = 1/N\delta t_o$.

Just what different filter functions we can generate depends on whether the mesh is a unitary mesh or a more general non-unitary mesh (e.g., as in Fig. S2). Quite generally, though, for some arbitrary desired filter function $H(\delta\omega) \equiv |H\rangle$, the "best" version of that filter we can generate (in a "least squares" sense), will be the one with the matrix elements

$$\tilde{M}_{qp} = \frac{1}{\sqrt{N}} \langle h_p | H \rangle \equiv \sqrt{\frac{2\pi}{N\delta t_o}} \int_{-\pi/\delta t_o}^{\pi/\delta t_o} \exp(ip\delta\omega\delta t_o) H(\delta\omega) d(\delta\omega) \qquad \text{(S55)}$$

This approach of finding the best representation of a function on a given basis by using the inner products with the basis functions as the expansion coefficients is standard in such linear algebra and functional analysis [27].

### Unitary filters and filter sets

With $M$ implemented as a unitary mesh, because of the construction of the multilayer unitary filter, the multiple filter functions it implements are physically guaranteed to be orthogonal to one another (at least if the mesh is "perfect" – that is, with 50:50 beam beamsplitters and the same loss on all paths through the mesh); each layer effectively implements a different row of a unitary matrix, and such rows are necessarily orthogonal. Any mutually orthogonal set of filter functions on the basis as in Eq. (S31) can be implemented. Because of the unitarity, the mesh overall will be loss-less; all the input power will appear somewhere in the $N$ outputs of the mesh itself (whether we use all $N$ of them or not).

If we express the filter functions as (normalized) vectors of amplitudes as in Eq. (S32), for example,

$$\left|H_q\right\rangle \equiv \begin{bmatrix} \tilde{M}_{q1} \\ \vdots \\ \tilde{M}_{qN} \end{bmatrix} \tag{S56}$$

any $Q$ ($\le N$) such orthogonal vectors can be implemented in the outputs of a mesh with $Q$ "layers" (such as the self-configuring layers as in Fig. 2 of the main text).

For example, instead of working directly on the Fourier-like basis of Eq. (S31), we could unitarily transform to and work with what we could call the "AWG" basis, which is functions each like those in Eqs. (S43) or (S44), but shifted (by effective mathematical phase tilts) so that the peak of a given filter function lines up with the zeros of every other such filter function. The actual filters that can be implemented do not depend on which such mathematical basis we choose, however.

### Non-unitary filters and filter sets

Though unitary meshes are restricted to orthogonal filter functions, for non-unitary meshes, as in Fig. S2, in principle we can program any desired set of filter functions on the same basis. For $N$ waveguides in the array, we can program up to $N$ different filter functions, and these need not be orthogonal; those filter functions, on the Fourier-like basis Eq. (S31), are then simply the rows of the (generally non-unitary) matrix $M$. We do, however, need to consider power conservation in the final set of filters that can be generated. Though it would be possible to add gain instead of (lossy) modulation in generating a non-unitary mesh, adding such gain can be technologically challenging, so we presume we will not do so. Given that constraint, a procedure for setting up the best mesh can be as follows:

1. Write down the initially desired matrix $M_{init}$ with the expansion coefficients of the desired filter functions as the rows.
2. Mathematically perform the SVD of $M_{init}$.
3. Note the required singular values, and in particular note the largest one, which, without loss of generality, we can call $s_1$.
4. Divide the matrix by $s_1$ (or, at least, its magnitude) to obtain $M_{actual} = M_{init} / |s_1|$. This guarantees that the largest singular value in $M_{actual}$ is no larger than 1 in magnitude, which means physically that we do not need to introduce gain in the system. (The SVD of this new matrix $M_{actual}$ is the same as that of $M_{init}$ except that all the singular values are divided by $|s_1|$). Such new singular values can be implemented by the line of modulators as in Fig. S2.

With this approach, we can implement the desired set of non-orthogonal filter functions, subject to their overall magnitudes being reduced, if necessary, by this factor $1/|s_1|$. This ability to generate multiple non-orthogonal filter functions simultaneously and physically in optics, and with independent programmability, is an unusual one.

### *Mesh architectures and topologies*

As we consider the full range of filters that could be made with this approach, we can usefully consider the topologies of architectures made from $2\times 2$ blocks such as the MZIs in Fig. 1 of the main text [28]. Quite generally, the architectures we are considering are all "forward-only"; there are no reflections or loops in the network. In filter terminology, because of the absence of loops or reflections, these filters are all "finite impulse response" or (non-autoregressive) "moving average" filters [8] (as opposed to "infinite impulse response" or "autoregressive" filters). In network topology nomenclature, these are "directed acyclic graphs". The $2\times 2$ (e.g., MZI) blocks can be viewed as the "nodes" in the graph and the waveguides between these nodes are the graph "edges".

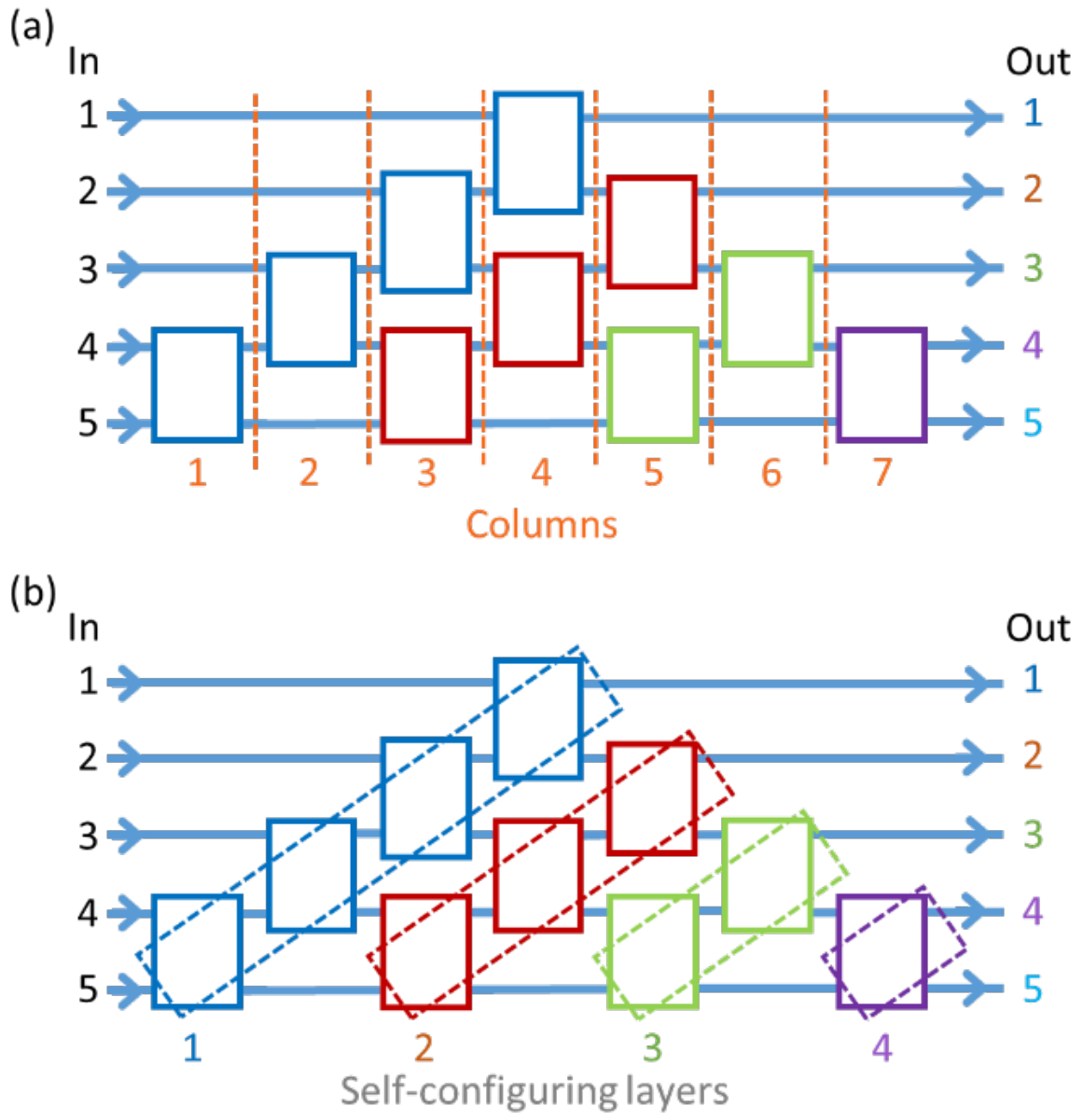

Fig. S3. Two important topologies for $2\times 2$ blocks (such as MZIs) in forward-only (directed acyclic graph) mesh networks, illustrated for a triangular mesh. (a) Column topology, in which all blocks in a column can be configured or calibrated in parallel; all forward-only networks can be arranged into this form. (The triangular network is unusual in that no waveguide crossings are required in such an arrangement.) (b) Self-configuring layer topology; in such a layer, each layer input connects to the "signal" output by only one path through the blocks in the layer. Not all forward-only networks have such topologies, though any network formed from cascades of successive self-configuring layers automatically does.

Any such directed acyclic graph can be arranged into "columns" without changing the topology. (As discussed in Ref. [28], this arrangement corresponds to Dijkstra's algorithm [29].) The blocks or nodes in a column do not affect the inputs to any other blocks or nodes in the same column, and so they can be calibrated or set in parallel. With rectangles representing the $2\times 2$ blocks, Fig. S3(a) shows the corresponding “column” topology in a “triangular” mesh [30] architecture.

Another important topology of some directed acyclic graphs or forward-only mesh architectures is the self-configuring layer topology [1] (Fig. S3(b)). Topologically, a self-configuring layer is one in which each input port is connected by only one path through the blocks to a single “signal output” port (or waveguide). The diagonal line and (symmetric) binary tree architectures [31] have this property, as do hybrids of them [1]. Conveniently, cascades of multiple such layers can be made. If we start with one such layer with $N$ inputs, we can essentially “extract” this signal output from that layer as a useful output. Then the other $N-1$ outputs (which we could call “drop-port” outputs) are passed as inputs to a succeeding self-configuring layer (which has its own “signal output” port). If we wish, we can continue this for a total of up to $N-1$ layers to form a full cascade.

Such a full cascade implements a fully programmable $N\times N$ unitary matrix. Fig. S3(b) shows self-configuring layers in a full cascade in a triangular mesh architecture, which we see can also be viewed as being formed from successive cascaded “diagonal line” self-configuring layer mesh architectures (as also in Fig. 1(c))). Not all mesh architectures can be factored into such successive self-configuring layers; notably, the rectangular architecture [30] cannot be factored this way.

Detectors can be added between columns or self-configuring layers to sample the amount of power in one or both outputs of a $2\times 2$ block, which can be convenient for calibrating or setting up the network. Such detectors can be “mostly transparent” detectors [32] in the waveguides, or can use waveguide “taps” to sample a small amount of power from the guide to a conventional detector [33]. Other techniques based on two output detectors [34] can also be used to deduce the power passing through any phase shifter in a forward-only mesh. With such monitoring of internal powers in the mesh, simple progressive techniques [28,35] allow mesh calibration by power maximization based on shining in specific vectors of amplitudes. Any given self-configuring layer can also be set up progressively just with a detector at its signal output; to configure the first $2\times 2$ block, we set up the other blocks to route this power to the signal output of that layer, and can proceed similarly through the subsequent $2\times 2$ blocks in the layer [31]; meshes based on cascaded self-configuring layers can also be set up progressively just with detectors at the outputs [1,19,31] in an extension of this approach.

### *Programming matrices on meshes*

Generally, the matrices corresponding to collections of $2\times 2$ blocks can be set up as matrix products of the $2\times 2$ matrices corresponding to each block. With a calibrated mesh, and with calculation of the factorization of the overall matrix into such $2\times 2$ blocks, arbitrary matrices can be set up this way. See, for example, Ref. [28] for a general discussion of this approach. Ref. [35] shows how to calibrate any forward-only mesh.

### Programming self-configuring layers

Architectures made from successive self-configuring layers are particularly easy to set up and to calibrate [1,19,31,35]. Explicit procedures for diagonal line and symmetric binary tree mesh layers are given in Ref. [1] and its supplementary material, including both simple calibration procedures and explicit formulas for mesh settings for a given vector. Self-configuring layers only require one input vector for each self-configuring layer to be calibrated or programmed. Shining in a vector $|\psi_1\rangle$ and optimizing progressively for minimum powers at the drop ports of each successive $2\times 2$ block configures the first such layer to have a matrix row of $\langle\psi_1|$. For each successive self-configuring layer $j$ in a unitary mesh, shining in a vector $|\psi_j\rangle$ that is orthogonal to all the preceding such vectors $|\psi_1\rangle, |\psi_2\rangle,\ldots,|\psi_{j-1}\rangle$ can similarly configure the succeeding matrix row to $\langle\psi_j|$, giving a simple way of programming any unitary matrix to the set of rows $\langle\psi_j|$ just by successively shining in the vectors $|\psi_j\rangle$ and configuring these successive layers.

It is also particularly straightforward to calculate the required settings of the self-configuring layer blocks to implement a desired matrix row. Conceptually, we can do this by imagining we are running the system backwards, shining light backwards into the signal output to generate a specific vector coming backwards out of the inputs to that mesh layer. Quite generally with unitary mesh networks (Ref. [1] supplementary section S5, Ref. [35]) if a given input vector $|\psi\rangle$ gives some output vector $|\phi\rangle$, then if instead we shine a vector of amplitudes $(|\phi\rangle)^*$ backwards into the "outputs", the vector that emerges from backwards from the inputs will be of the form $(|\psi\rangle)^*$. (When we say "of the form", we mean that the relative amplitudes and phase delays are the same as those of the vector $(|\psi\rangle)^*$. If there is any overall common loss in the forward direction, there will be the same loss in the backwards direction, so the overall amplitude of the backwards vector will be reduced accordingly.) This is a general property of reciprocal loss-less optical systems, and relates to phase conjugation (see Ref. [1] supplementary material for an explicit discussion of these phase conjugating properties in mesh mathematics); $(|\phi\rangle)^*$, which is the column vector whose elements are the complex conjugates of those in the vector $|\phi\rangle$, can be called the phase conjugate of $|\phi\rangle$.

Now, the phase conjugate of a beam going "forwards" in a single-mode guide is just the same mode going "backwards". If we supposed that the layer is set up to take all of the power in vector $|\psi\rangle$ and put it in the signal output port of that layer, then just shining light backwards into that signal output port (so "phase conjugating" that simple output) would therefore generate the vector $(|\psi\rangle)^*$ coming backwards out of the mesh "inputs". So, to design the settings of the self-configuring layer to take all of the power in input vector $|\psi\rangle$ and put it at its signal output, we simply imagine that we are shining light back into the signal output and calculate how to generate the vector $(|\psi\rangle)^*$ emerging backwards from the "inputs". For the diagonal line mesh, this is particularly obvious.

Using the diagonal mesh of Fig. 1(c), we set up the $\theta$ phase shifter of MZI3 in Fig. 1(c), to split out the desired fraction of power into WG1, and then set the $\phi$ phase shifter of MZI3 to give the desired phase. Then we set up the $\theta$ phase shifter of MZI2 to split out the desired fraction of the remaining power into WG2 and set its phase with the $\phi$ of MZI2, and so on. For other self-configuring layer meshes such as the symmetric binary tree, the calculations are slightly different but are still similarly progressive and straightforward.

In calculating how to set up successive self-configuring layers, we simply note the unitary matrices we have set up in earlier layers and compensate for those. So, for example, having set up the first layer in the mesh to correspond to the first matrix row $\langle\psi_1|$, which will correspond to a unitary matrix $U_1$ as in Fig. 2 of the main text, then to calculate how to set up the second row so it implements the row $\langle\psi_2|$, we set it up to send all the power in the vector $U_1|\psi_2\rangle$ to its signal output, and so on for each successive layer, multiplying all the unitary matrices corresponding to all previous layers as necessary. This algebra is explicitly given in [19].

Calculating how to set up the general non-unitary matrix in the SVD configuration is also straightforward with unitary meshes made from successive self-configuring layers; each of the two unitary meshes can be set up (or even self-configured) in the same way.

### Simultaneous self-configuration with identifying tones

In an input signal comprised of multiple different optical frequencies, by putting different identifying tones on different specific optical frequencies in that input signal, such as some small amplitude modulation at a given tone frequency, layers in the network can be self-configured to extract such specific optical frequencies to specific signal outputs. For example, the detection process for maximizing or minimizing power, as required for self-configuration in a given layer, can also incorporate electronic filters that have the system look only for such specific tones and the amplitude of the corresponding modulation in the detected signals. (The use of such modulation tones or frequencies for setting up self-configuring layers is discussed in [31] and used, for example, in Ref. [32].) Such a process then can optimize that layer to extract, to its signal output, signals on this specific optical frequency. (This extraction of multiple frequencies by multiple layers can only be perfect if the corresponding filter functions are orthogonal, of course.)

## S4. Analyzing temporal partial coherence with self-configuring filters

Here we give a formal discussion of the use of the device concept to analyze temporal partial coherence. We use a modal description of partial temporal coherence, using approaches from references such as Wolf [36] and others [37]. We consider the amplitude of the field mode in the single-mode input waveguide to be a single time-varying scalar field function $x(t)$. We formally presume that this function of time is one of a statistical ensemble $x_j(t)$ of such possible functions, where *j* indexes all the possible functions in the ensemble. Quite generally, the (first-order) coherence (or coherence function) of such an ensemble between the values of the function at two times $t_1$ and $t_2$ can be written as

$$\Gamma(t_1,t_2) = \left\langle x^*(t_1)x(t_2)\right\rangle_j \tag{S57}$$

where by $\langle \cdot \rangle_j$ we mean the average over this statistical ensemble. In our case, we choose to write

$$t_1 = t + u\delta t_o \, , \; t_2 = t + v\delta t_o \tag{S58}$$

where *t* is time, $\delta t_o$ is our time delay element characterizing the waveguide array (as in Eq. (S16)), and for the moment *u* and *v* are real numbers (though later those will be integers characterizing waveguide relative lengths).

We take these $x_j(t)$ to be ergodic random variables, by which we mean that an average over this ensemble of possibilities at some given time is the same as a time average. This in turn requires we presume the process is also "stationary", with probabilities independent of the chosen time origin. This choice means we presume that (average) power measurements we might make on these fields will be independent of time. With these ergodic and stationary assumptions, the average over time will be the same as the ensemble average. We can also freely shift the time origin for the averaging because of the stationarity property, and any such coherence function is then only a function of time differences. As a result, we can define the mutual coherence function $\Gamma(\tau)$ of the field as [36] (Eq. (8) of the main text)

$$\Gamma(\tau) = \left\langle x^*(t)x(t+\tau)\right\rangle_t \tag{S59}$$

where $\langle \cdot \rangle_t$ denotes time averaging, and where

$$\tau \equiv (v-u)\delta t_o \tag{S60}$$

This time averaging can be expressed as the integral

$$\Gamma(\tau)=\lim_{T\to\infty}\frac{1}{2T}\int_{-T}^{T}x^{*}(t)x(t+\tau)\,dt \tag{S61}$$

Note, incidentally, because this is presumed to be a stationary process, shifting the time origin again by $\tau$, makes no difference to the averaging. Hence

$$\Gamma(\tau)=\lim_{T\to\infty}\frac{1}{2T}\int_{-T}^{T}x^{*}(t-\tau)x(t)\,dt=\left[\lim_{T\to\infty}\frac{1}{2T}\int_{-T}^{T}x^{*}(t)x(t-\tau)\,dt\right]^{*}=\Gamma^{*}(-\tau) \tag{S62}$$

Consider now the spectrometer system from Fig. 2 of the main text, with the "AWG-like" set of waveguides with lengths spaced progressively by increments $\delta l_o$. In this case, it is more useful to think of the waveguides as giving a set of time-delayed versions of the input signal. So, we have progressive relative time delays in waveguide *p* of $p\delta t_o$, with $\delta t_o = n_g \delta l_o / c$ (Eq. (S16), also Eq. (1) of the main text), and with equal power splitting, so a factor $a_p = 1/\sqrt{N}$ as in Eq. (S1) above. After passing through the power splitter and waveguide array, the input to the mesh *M* is a set of time-delayed replicas of the signal $x(t)$ of $(1/\sqrt{N})x(t-p\delta t_o)$ in the *p*th input waveguide. As a result, the output field in waveguide *q* is

$$y_q(t)=\frac{1}{\sqrt{N}}\sum_{p=1}^{N}M_{qp}\,x(t-p\delta t_o) \tag{S63}$$

(We could also view this formally as the result of convolution of the input signal $x(t)$ with the impulse response of the system in waveguide *q*, as in Eq. (S35).)

We can formally write

$$y_q^{*}(t)=\frac{1}{\sqrt{N}}\sum_{s=1}^{N}M_{qs}^{*}\,x^{*}(t-s\delta t_o)\equiv\frac{1}{\sqrt{N}}\sum_{s=1}^{N}M_{sq}^{\dagger}\,x^{*}(t-s\delta t_o) \tag{S64}$$

So, the time-averaged power in waveguide *q* is

$$\begin{aligned}
\left\langle |y_q(t)|^2\right\rangle_t &\equiv \left\langle y_q(t)y_q^{*}(t)\right\rangle_t=\left\langle \frac{1}{N}\sum_{p=1}^{N}M_{qp}\,x(t-p\delta t_o)\sum_{s=1}^{N}M_{sq}^{\dagger}\,x^{*}(t-s\delta t_o)\right\rangle_t \\
&=\left\langle \frac{1}{N}\sum_{p=1}^{N}\sum_{s=1}^{N}M_{qp}\,x^{*}(t-s\delta t_o)x(t-p\delta t_o)M_{sq}^{\dagger}\right\rangle_t \\
&=\frac{1}{N}\sum_{p=1}^{N}\sum_{s=1}^{N}M_{qp}\left\langle x^{*}(t-s\delta t_o)x(t-p\delta t_o)\right\rangle_t M_{sq}^{\dagger} \\
&=\frac{1}{N}\sum_{p=1}^{N}\sum_{s=1}^{N}M_{qp}\left\langle x^{*}(t)x(t+(s-p)\delta t_o)\right\rangle_t M_{sq}^{\dagger} \\
&=\frac{1}{N}\sum_{p=1}^{N}\sum_{s=1}^{N}M_{qp}\,\Gamma((s-p)\delta t_o)M_{sq}^{\dagger}\equiv\frac{1}{N}\sum_{p=1}^{N}\sum_{s=1}^{N}M_{qp}\,\Gamma_{ps}M_{sq}^{\dagger} \\
&\equiv\frac{1}{N}\left(M\Gamma M^{\dagger}\right)_{qq}
\end{aligned} \tag{S65}$$

where what we will now call the temporal coherency matrix $\Gamma$ in the last line has the matrix elements (Eq. (9) of the main text)

$$\Gamma_{ps}\equiv\Gamma((s-p)\delta t_o) \tag{S66}$$

These matrix elements therefore represent a discretized version of the mutual coherence function as in Eq. (S59). Note that, by Eq. (S62),

$$\Gamma_{ps} = \Gamma_{sp}^{*} \tag{S67}$$

So, this matrix $\Gamma$ is Hermitian, which means it has real eigenvalues and orthogonal eigenvectors. We can therefore choose the rows of the unitary matrix $M$ to correspond to the eigenvectors of $\Gamma$. In that case, the matrix $M\Gamma M^{\dagger}$ is diagonal, and, for unit input power, the output powers as given by Eq. (S65) are the eigenvalues of $\Gamma$. Because the left hand side of Eq. (S65) is the average of a positive number, we can see directly that all the eigenvalues of $\Gamma$ are necessarily non-negative.

Now, in the presence of the input field, if we adjust the first layer of the mesh to maximize its output power, then we are variationally setting the first row of the matrix to correspond to the eigenvector with the largest eigenvalue. Leaving those settings, we can then similarly maximize the power from the second mesh layer, which will set the combination of the first and second layers to correspond to the second eigenvector and so on. In this way, we can establish the eigenvectors and eigenvalues of $\Gamma$, hence measuring this discretized version of the coherency matrix or coherence function. (This is the same sequential mesh optimization procedure introduced by some of us in the context of measuring the spatial coherency matrix [38].)

Note too that the input field has then been separated by the mesh into its eigen components, which in this case correspond to the "natural" mode or Karhunen-Loève expansion, providing the separate field at the signal outputs of the mesh. These output powers will also be mutually completely incoherent. We are not aware of another approach that accomplishes both this measurement and this non-destructive physical separation, which is simultaneous for all these components and is without any loss (other than background losses in the system).

Incidentally, having set up the matrix $M$ in this way by physically separating into its natural mode components, it is now straightforward to understand exactly what set of filter functions we have constructed. We know the matrix elements $M_{qp}$, and so we can simply deduce the filter functions from Eq. (S20) (Eq. (3) of the main text).

Above we have presumed for simplicity in the derivation that the waveguides in the array have lengths $p\delta t_o$ where $p$ is the summation index, so $p$ takes on sequential integer values. We note that, as a result, several matrix elements of $\Gamma_{ps}$ are identical (e.g., $\Gamma_{12} = \Gamma_{23}$), so we are essentially measuring the same value of the coherence function multiple times. If instead we chose sets of relative lengths $m_p\delta t_o$ such that all the differences between the various $m_p$ values corresponded to different integers, then we could avoid this redundancy, and obtain measured values of $\Gamma(\tau)$ at a larger number of different time-delay values $\tau$ for the same number of physical measurements in the apparatus. Of course, this simply corresponds to making the set of integers $m_p$ form a non-redundant set, as in a Golomb ruler, and as discussed for the non-redundant array filters in the main text. So, using such a non-redundant array approach for the waveguides gives us a more efficient method of measuring the coherence function for a larger number of different time (delay) values.

## S5. Other operating modes and extensions

Here we give an extended discussion of some additional device operating modes and extensions to the device architecture.

### *Running the device backwards*

Though we have so far considered the device working forwards from the input waveguide to one or more output waveguides, as in the architectures of Fig. 1 or 2, there are several possibly useful ways of running the device backwards.

One example would be wavelength multiplexing. We could, for example, shine light of a first wavelength back into the signal output 1 of Fig. 2. We could then adjust the settings of that first mesh layer to maximize the power coming back out of the input waveguide of the entire device. This would be equivalent to lining up the peak of the response as in Fig. 3 to coincide with that wavelength. Then, leaving those settings, we could now shine light of a second wavelength back into signal output 2 of Fig. 2, and adjust the settings of the second layer to again maximize the power coming back out of the input waveguide. If that second wavelength happened to line up with one of zeros of the filter response we have effectively set on the first layer, then we should be able to couple all of this second wavelength backwards out of the input waveguide. To the extent that the second wavelength does not line up perfectly with such a zero, there will be some loss in multiplexing this second wavelength (which would appear as loss in the power splitter, scattering light out of it). Nonetheless, even with some loss from such imperfect spacing of the wavelengths to be combined, we should still be able to couple much of the power from this second wavelength back out of the input guide. We can continue to extend this for different wavelengths backwards into the outputs of other layers in Fig. 2. A feature of this multiplexer is that it will still multiplex wavelengths together (with some loss) even if they are not quite equally spaced, and without knowing in advance just what the wavelengths are.

A second example of running the device backwards would be to operate the device as a wavelength-dependent mirror by adding an actual mirror or reflector at the signal output of a layer. Such a configuration used as a mirror in a laser cavity could allow tuning of the laser, for example over the free-spectral range of the device, with the laser running at or near the wavelength corresponding to the peak power transmission to that signal output.

A third example would be to use the device as a pulse generator. If the various outputs of a mesh as in Fig. 2 were each fed backwards by different mutually coherent elements of a frequency comb, the mesh could combine them controllably to generate pulses at a repletion rate given by the difference frequency between adjacent frequencies in the comb. Generally, we can imagine various architectures related to those of the schemes as reviewed in Weiner et al. [39] for pulse generation and shaping, with architectures such as those of Fig. 2 replacing grating dispersive elements, for example, and allowing quite arbitrary pulse shape control.

### *Other modalities exploiting partial coherence*

#### Separating high-coherence sources or spectral lines

If the input spectrum contains some narrow lines amid a broad, incoherent background, the approach above proposed for measuring the coherence function will tend to separate out such coherent lines preferentially in the first layer or layers of the mesh because those will tend to be the strongest components that are coherent with themselves and mutually incoherent with respect to the rest of the spectrum. Note that this will happen without prior knowledge of the frequency of any such spectral lines.

#### Configuring for absorption lines rather than spectral emission lines

The discussion so far has largely concerned the detection and separation of specific frequencies or wavelengths of light to different outputs. This is also appropriate if we are performing spectroscopy where we are looking for different spectral lines in the emission from gases of atoms or molecules, for example. Often, however, spectroscopy is concerned with looking for corresponding absorption lines, again, for example, in gases of atoms or molecules. In such a case, the material to be investigated, such as a gas, may be illuminated with a broad, incoherent spectrum of light, such as from the sun or some light bulb. To look for absorption lines, one can then configure or tune the first mesh layer to look for minimum rather than the maximum power in its signal output. Configuring a second layer to look for a maximum output power would tend to mean that the second layer was effectively measuring the background light power in regions with minimal absorption. The amount of absorption from the spectral line of interest can then be deduced by comparing the output of this first layer to the output of this second layer. Alternatively, if we want to compare the relative strength of two such absorption lines, so we can tell the relative abundance of two species, for example, we can compare the minimum power from two layers, one tuned to the first spectral line and the second tuned to the second such line.

#### Separating channels without knowing the wavelengths

Signals from different lasers can generally be considered mutually incoherent. So, for example, if we have several telecommunications signals from different lasers incident at the same time in the input, a mesh such as that in Fig. 2 could at least partially separate them automatically just based on their mutual incoherence as discussed above. In such a case, we might decide to optimize each layer based not on maximum output power, but instead on, say, the degree of opening of an eye diagram or a minimum in a bit error rate (see also the discussion below on optimization on other parameters). If the signals from the different lasers are of somewhat different powers and if their wavelengths are separated by more than the basic spectral resolution of the system (as given by the widths of the "peaks" in Fig. 3, for example), then we can expect that these signals can be usefully separated, at least approximately. Note that such a separation does not require advance knowledge of the specific wavelengths of the signals.

### *Extensions to the device architecture*

#### NxQ meshes

Working with a large number $N$ of input waveguides together with a smaller number $Q$ of self-configuring layers can allow a relatively compact mesh (in terms of the number of MZIs required) that nonetheless has good spectral resolution and discrimination between different signals or spectra. Such an architecture could, for example, be programmed to pick out just a few ($Q$) spectral lines in a larger spectrum, presenting their powers at the $Q$ output waveguides. It could also be a good mesh for separating a few wavelength channels when the input wavelengths are not known, as discussed above, because, with a large $N$, it more easily discriminates closely spaced channels.

#### Adding a controllable power splitter

Instead of the fixed power splitter of Figs. 1 and 2, we could instead use a controllable splitter, similar to the diagonal line or binary tree networks as shown in Fig. 1 but run backwards (i.e., with the input light shining "backwards" into the signal "output" of the layer). Such a device could allow us to pre-compensate for different losses in different waveguides so that the powers leaving the waveguides as they couple into the mesh are all approximately equal; this might be particularly useful for the non-redundant array architectures where waveguide lengths can be very different. In such a device, only controllable power splitting would be necessary (so only the $\theta$ phase shifters in the MZIs) because phase differences between guides are already handled by the mesh $M$.

#### Successive layers or filters for better rejection

With ideal components, a given layer can completely reject a given wavelength. Of course, imperfections exist; rejection will not generally be perfect, and some of that wavelength will also be passed to the next layer. For better rejection, one simple strategy is then to configure the second layer also to reject the same wavelength. If there is no other leakage of light in the system, we can expect this second layer to lead to a doubling of the overall rejection (in decibels or some other logarithmic scale).

Of course, as with any filtering system, we can cascade complete filters, here also including another waveguide array, to improve rejection. Such a technique could be particularly effective with non-redundant array filters because the second set of waveguides could use a different set of lengths so their spectral "leakage" would not line up with the leakage in the first filter.

### Optimization on other parameters

Instead of optimizing based on maximizing or minimizing power out of the signal port or ports, the device could be run to optimize other measurable parameters. For example, it could optimize based on "eye-opening" in telecommunications eye-diagrams or on minimum bit error rate, which essentially is a technique to correct for pulse dispersion and/or crosstalk with other channels. A related concept is to maximize the measured signal-to-noise ratio, which tends to create matched filters [40,41]. A single-layer filter could compensate and extract the single best channel, and multilayer filters could extract or compensate additional channels.

### Handling imperfections and perfect MZIs

One of the possible imperfections in MZIs is that the beamsplitter split ratio will not in general be 50:50. That imperfection can limit rejection, for example, especially in "diagonal line" architectures. (Given approximately equal desired power splitting overall in a diagonal line, the overall power splitting of the individual MZIs in a long line tends to vary from nearly unity at one end of the line to nearly zero at the other. Symmetric binary trees are less susceptible to this issue since the corresponding power splitting of all MZIs tends to be near 0.5 for each MZI [42].)

One solution to this issue is to use "double" or "perfect" MZIs in which the actual beamsplitters are themselves implemented by "beamsplitter" MZIs. Such a "beamsplitter" MZI can behave as a 50:50 beamsplitter even for fabricated split ratios as bad as 85:15 [43]. There is also an algorithm that allows these MZIs to be set to 50:50 effective split ratio based only on successive power minimizations or maximizations [43], and high rejection has been reported in such configurations [44]. All such "perfect" MZIs inside a larger mesh can be set up with a sequence of such power minimization or maximization algorithms [43].

### Dithering for derivative spectra

Derivative techniques can be generally useful in spectroscopy; they can identify line centers precisely, for example, as those correspond to zeros in the first derivative, and overall background light can be easily rejected. One way to perform such derivative spectroscopy here would be to add some small modulation of the $\gamma$ tuning parameter, at some frequency $\nu$ and look for modulation at that frequency for first-derivative spectra (or at $2\nu$ for second-derivative spectra) at the signal output of a given layer. This could be achieved by adding small modulations of amplitude $\propto m_p$ (as in Eq. (7) of the main text) to the "input" phase shifters (e.g., PS1 to PS4 in Fig. 1). A particularly simple way of doing that would be to modulate the temperature of the entire waveguide array at such a frequency $\nu$ since that would add phase shift appropriately proportional to the waveguide length (i.e., $\propto m_p$ ) in every waveguide in the array.

### Spatio-spectral device

The spectral mesh concept discussed here could also be combined with spatial meshes. For example, if we had, say, $W$ different input waveguides, with their inputs coming from different grating couplers or other spatial mode couplers, we could split each of those into its own array of $N$ waveguides of different lengths. We could then feed all those waveguides into a mesh with $W \times N$ inputs. That mesh could then implement simultaneous spatial and spectral operations. For example, it could possibly extract different spatial modes from a multimode fiber and perform different dispersion-compensating spectral filtering on different modes. It could also operate simultaneously on spatial and spectral partial coherence. Such a concept could also be extended to using non-unitary meshes (e.g., in the SVD architecture [19]).